\documentclass[letterpaper,12pt]{article}
\usepackage[margin=.75in]{geometry}
\usepackage{authblk}
\usepackage{hyperref}
\usepackage[superscript]{cite}
\usepackage{graphicx}
\usepackage[labelfont=bf]{caption}
\usepackage{amsmath,amssymb,amsfonts}

\usepackage{titlesec} 
	\titleformat{\section}{\Large\bfseries}{}{0pt}{}
    \titleformat{\subsection}{\large\bfseries}{}{0pt}{}
	\titleformat{\subsubsection}{\bfseries}{}{0pt}{}

\usepackage{booktabs}%

\usepackage{multirow}%
\usepackage{rotating}
\usepackage{array}

\DeclareUnicodeCharacter{03B5}{$\epsilon$}
\DeclareUnicodeCharacter{2217}{*}
\DeclareUnicodeCharacter{202F}{\,}

\hypersetup{hypertexnames=false,colorlinks,allcolors=black}
\usepackage{lineno}
\usepackage{setspace}

\newcommand{\arar}{\texorpdfstring{\textsuperscript{40}Ar-\textsuperscript{39}Ar}{40Ar-39Ar}}
\newcommand{\kar}{\texorpdfstring{\textsuperscript{40}K-\textsuperscript{40}Ar}{40K-40Ar}}
\newcommand{\iso}[2]{\textsuperscript{#1}#2}
\newcommand{\K}[1]{\textsuperscript{#1}K}
\newcommand{\Ar}[1]{\textsuperscript{#1}Ar}
\newcommand{\lognorm}[1]{$log\mathcal{N}(#1)$}
\newcommand{\unf}[1]{$\mathcal{U}[#1]$}
\newcommand{\yss}{y\textsubscript{ss}}
\newcommand{\ImpactChron}{\texttt{ImpactChron.jl}}

\begin{document}
\onehalfspacing

\title{An early giant planet instability recorded in asteroidal meteorites}

\author[*~1,2,3]{Graham Harper Edwards}
\author[3]{C.~Brenhin Keller}
\author[2,3]{Elisabeth~R.~Newton}
\author[3]{Cameron~W.~Stewart}

\affil[1]{Earth \& Environmental Geosciences, Trinity University, 1 Trinity Place, San~Antonio, TX 78212 U.S.A.}
\affil[2]{Physics \& Astronomy, Dartmouth College, 17 Fayerweather Hill Road, Hanover, NH 03755 U.S.A.}   
\affil[3]{Earth Sciences, Dartmouth College, 19 Fayerweather Hill Road, Hanover, NH 03755 U.S.A.}

\date{}\maketitle

\textbf{*Corresponding author's email:} graham.h.edwards@dartmouth.edu


\abstract{
Giant planet migration appears widespread among planetary systems in our Galaxy.
However, the timescales of this process, which reflect the underlying dynamical mechanisms, are not well constrained, even within the solar system. 
Since planetary migration scatters smaller bodies onto intersecting orbits, it would have resulted in an epoch of enhanced bombardment in the solar system's asteroid belt. 
To accurately and precisely quantify the timescales of migration, we interrogate thermochronologic data from asteroidal meteorites, which record the thermal imprint of energetic collisions. 
We present a database of \kar{} system ages from chondrite meteorites and evaluate it with an asteroid-scale thermal code coupled to a Markov chain Monte Carlo inversion. 
Simulations require bombardment in order to reproduce the observed age distribution and identify a bombardment event beginning $11.3^{+9.5}_{-6.6}$~million years after the Sun formed (50\% credible interval). 
Our results associate a giant planet instability in our solar system with the dissipation of the gaseous protoplanetary disk. 
}

\section{Introduction}\label{intro}

Planetary migrations seem to be commonplace in our  Galaxy.
The proximity of ``hot Jupiters'' to their host stars results from inward migration from more distant planetary birth radii~\cite{Dawson2018}. Planets in the TRAPPIST-1 system also likely migrated inward from larger radii where they inherited their volatile inventories~\cite{Unterborn2018inward}.
Distributions of both exoplanet eccentricity and orbital spacing in multi-planet systems are most readily explained by histories of dynamical instability and orbital reorganization (e.g.~refs.\cite{Pu2015,Raymond2010planetplanet}).

Several lines of evidence indicate that the solar system's giant planets underwent at least one episode of migration. 
The admixture of material from the inner and outer solar system among main belt asteroids~\cite{DeMeo2014solar} and asteroidal meteorites~\cite{Kruijer2020great} requires dynamical mixing of protoplanetary reservoirs.
The orbital architecture of giant planets~\cite{Fernandez1984dynamical,Tsiganis2005} and the Kuiper Belt~\cite{Gomes2003origin} as well as the low masses of Mars and the asteroid belt  ~\cite{Walsh2011,Clement2018mars,Clement2019excitation} could not have formed in situ and require a history of dynamical excitation. 
Consequently, giant planet migration (GPM) established the long-term ($>$4~billion-year) physical and chemical structure of the solar system \cite{DeMeo2014solar,Kruijer2020great,Fernandez1984dynamical,Tsiganis2005,Gomes2003origin,Walsh2011,Clement2018mars,Clement2019excitation} and perhaps promoted terrestrial habitability by supplying volatile-rich material from the outer solar system to the early Earth~\cite{Kasting2003evolution,Alexander2012provenances,Fritz2014}. 
As a corollary, we expect migrations to similarly imprint these characteristics in exoplanetary systems~\cite{Fritz2014}.

\begin{figure}
    \centering
    \includegraphics[width=\textwidth, keepaspectratio]{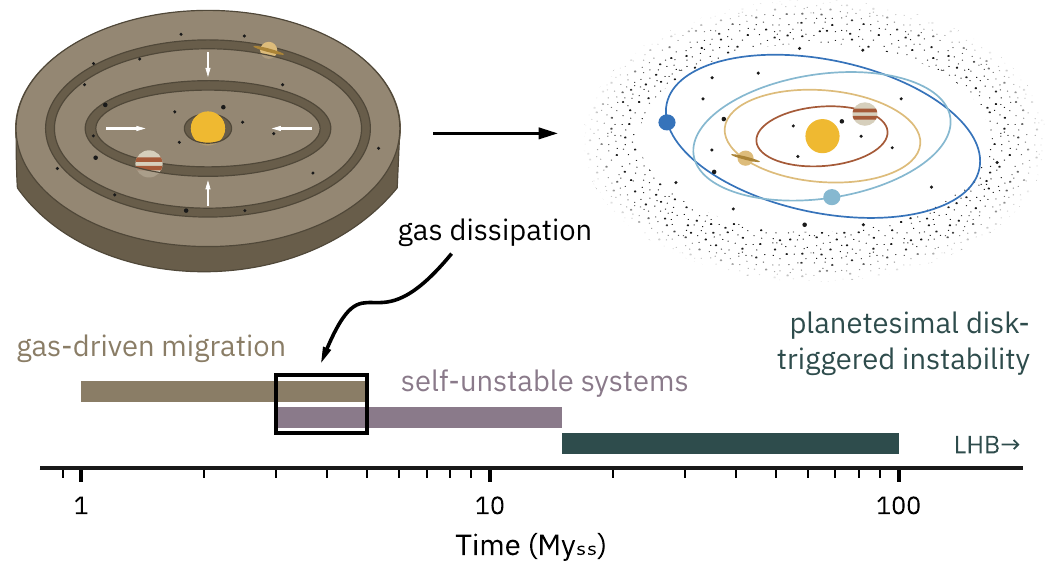}
    \caption{
    \textbf{Timescales of giant planet migration stimuli.}
    Diagram on the left depicts inward (Type II) migration of a proto-Jupiter and proto-Saturn within gaps in the gaseous protoplanetary disk (in brown). 
    The solar system's gaseous disk dissipated $\le$5~M\yss{}~\cite{Borlina2022lifetime}. 
    Diagram on the right depicts a giant planet instability (giant planets on high-eccentricity orbits) in the presence of an outer planetesimal disk. 
    Such an instability might be caused by dissipation of the gaseous disk, a giant planet orbital configuration that is inherently unstable without the presence of the gaseous disk, or interactions with an outer planetesimal disk.
    Gas dissipation occurs between 3--5~M\yss{} (black box), destabilization of a self-unstable system occurs $\le$10~My after gas dissipation ($\sim$5--15~M\yss{}), and planetesimal disk-triggered instability occurs within 100~My of gas dissipation~\cite{RibeirodeSousa2020}. 
    A Late Heavy Bombardment (LHB) scenario occurs $>$400~M\yss{}, beyond the scale of the timeline. 
    Note that each scenario corresponds to a distinct timeframe in solar system history, and the resultant dynamical excitation scatters inner solar system bodies (black dots). 
    M\yss{} reflects million years after solar system formation, assigned to the age of the oldest Ca-Al-rich inclusions.}
    \label{fig:timescales}
\end{figure}

{While dynamical models of GPM vary widely in their assumptions and details, the overarching mechanisms} that drive GPM fall into one of two categories (Fig.~\ref{fig:timescales}): dynamical instability triggered by interplanetary gravitational interactions~\cite{Fernandez1984dynamical,Gomes2003origin,Tsiganis2005,Clement2018mars, RibeirodeSousa2020} or inward migration triggered by tidal interactions with a surrounding gaseous disk, also known as ``Type II'' migration~\cite{Lin1986tidal,Kley2012planetdisk}.
Since these two processes respectively require the absence or presence of a gaseous protoplanetary disk and gaseous disks are transient features~\cite{Williams2011protoplanetary}, constraining the timescale of migration may help resolve its underlying stimulus (Fig.~\ref{fig:timescales}). 
We examine the solar system as a natural laboratory to test the timescales of GPM and identify the mechanism that best describes solar system history and chronology.

A dynamical instability was first proposed as an explanation for the hypothesized Late Heavy Bombardment (LHB, described below)~\cite{Gomes2005origin} and the orbital architecture of the outer solar system \cite{Fernandez1984dynamical,Gomes2003origin,Tsiganis2005}. 
In these original giant planet instability (hereafter ``instability'') models, interactions between the giant planets and an outer planetesimal disk caused secular migration of Jupiter and Saturn into a 1:2 mean motion orbital resonance. 
This excited the planets' orbits and caused a solar-system scale instability~\cite{Tsiganis2005}: a period of chaotic and intersecting planetary orbits.
More recent simulations have revealed a variety of plausible instability stimuli~\cite{RibeirodeSousa2020,Liu2022early}.

A gas-driven migration model was originally proposed to explain the low mass of Mars and the admixture of bodies from the inner and outer solar system in the asteroid belt~\cite{Walsh2011}.
In this model, the proto-Jovian and proto-Saturnian cores formed embedded in a gaseous protoplanetary disk.
As they reached sufficiently large masses, they carved out gaps in the disk and migrated inwards due to associated torques, following the pattern of Type~II planet migration~\cite{Lin1986tidal,Kley2012planetdisk}. 
Jupiter began migrating before Saturn, until the latter began a faster migration.
When the cores reached a 2:3 mean motion resonance, their migration reversed outward---a so-called ``Grand Tack''---until the disk eventually dissipated and froze their orbital positions.

In all GPM scenarios, migrating planets scatter smaller bodies onto dynamically excited and intersecting orbits, resulting in a surge of collisions \cite{Gomes2005origin,Walsh2011,Clement2018mars,Clement2019excitation}.
Conversely, acute collisional episodes require dynamical excitation, and we are not aware of a hypothesis other than GPM that so reliably accomplishes widespread collisions.
Thus, bombardment events seem to be a reliable proxy for GPM.
We posit that the epoch of enhanced collisions resulting from GPM is a diagnostic event that may be recorded in planetary records~\cite{Gomes2005origin,Walsh2011}, and its timescales, if precisely constrained, could be used to identify the causal dynamical processes (Fig.~\ref{fig:timescales}). 

We reconstruct the timescales of reheating during GPM-triggered bombardment by evaluating the thermochronologic records of asteroidal meteorites. 
Thermochronologic mineral systems record the timescales of thermal processes through the temperature-dependent retention of radiogenic isotopes in mineral lattices.
As asteroids cool to sufficiently low temperatures, these minerals retain radiogenic daughter isotopes and record a ``cooling age.''
The energy released by collisional impacts are well-established sources of heat in (proto)planetary systems and are recorded in their thermochronologic records (e.g.~refs.\cite{Bogard1995impact,Tera1974isotopic}).

This study integrates early solar system chronologies from a variety of sources, including cosmochronologic data and physics-based simulations, which rely on fundamentally different timescales. 
Several cosmochronologies (e.g.~\kar{} ages) are derived from radioisotopic measurements that record time relative to the present. 
Following the conventions of isotope geo- and cosmochemistry, we report these ages in \textit{annum} (a), meaning ``years before the present''. 
However, physics-based simulations record time after reference events.  
To connect cosmochemical records with astrophysical processes, a solar time-zero is canonically anchored to Ca-Al-rich inclusions (CAIs), the solar system's oldest macroscopic solids, which condensed from a solar composition gas\cite{Grossman1972condensation} heated by the infant Sun as it entered its pre-main-sequence phase\cite{Brennecka2020astronomical}.
We report astronomically referenced time in terms of years after CAIs (\yss), such that $0~\text{M\yss{}} \equiv 4567.3~\text{Ma}$~\cite{Connelly2012absolute}, though our results are insensitive to the chosen datum (Methods).

Prior work on the solar system's GPM history has relied on a combination of thermochronologic data and the modern solar system architecture.
The LHB was first hypothesized as a period of enhanced lunar bombardment at $\sim$4~Ga, based on the absence of $>$4~Ga thermochronologic cooling ages in Apollo mission return samples and an apparent paucity of $>$4~Ga craters on the lunar surface \cite{Wetherill1975,Tera1974isotopic}.
Subsequent studies revealed that neither of these lunar records actually required a $\sim$4~Ga lunar bombardment~\cite{Fassett2012lunar,Boehnke2016} and could be explained instead by a monotonic decline in bombardment flux since the epoch of planetary assembly. 
More recently, dynamical constraints on instability triggers \cite{RibeirodeSousa2020}, the survival of binary asteroids \cite{Nesvorny2018evidence}, and meteorite thermochronology \cite{Mojzsis2019onset} have constrained the timescales of GPM to the solar system's first 100~My.

A variety of models for ``early'' ($<$100~M\yss{}) GPM fall within different timeframes (Fig.~\ref{fig:timescales}).
{While each model varies in its physical details and relies on its own suite of assumptions, the fundamental stimuli for GPM respectively rely on the presence, absence, or dissipation of the gaseous protoplanetary disk.}
The Grand Tack model requires the presence of a gaseous protoplanetary disk~\cite{Walsh2011}.
Observations of solar-mass stars limit gas-disks to the first 10~My of stellar lifetimes~\cite{Haisch2001disk,Sung2009spitzer,Williams2011protoplanetary}, with a typical lifetime of $<$5~M\yss{} derived from observations~\cite{Li2016lifetimes}, dynamical models~\cite{Kimura2016birth}, and cosmochemical constraints~\cite{Borlina2022lifetime}.
In contrast, instability models require the dissipation of the gaseous disk during or prior to the GPM event.
In the earliest scenario, instability occurs during gas dissipation, due to asymmetric torques on giant planets from the inner edge of the outward-migrating gas disk~\cite{Liu2022early}.
After gas dissipation, simulations seeded with plausible planet formation/evolution histories~\cite{Izidoro2015accretion} reveal that an early giant planet instability can occur by canonical planet-planetesimal disk interactions or by a self-triggered instability resulting from unstable orbital architectures established while embedded in the gas-disk~\cite{RibeirodeSousa2020}.
The self-triggered instabilities occur rapidly, within 10~My after gas dissipation ($\sim$15~M\yss{}), whereas self-stable systems interacting with a planetesimal disk typically experienced instabilities $>$30~My after gas dissipation.
In summary, plausible GPM stimuli occur at $\lesssim$5~M\yss{} (gas-disk stimulus or gas dissipation), 5--15~M\yss{} (unstable orbital architectures), or $>$15~M\yss{} (planetesimal-disk interactions) (Fig.~\ref{fig:timescales}).

This study aims to resolve the timescale of GPM and concomitant inner solar system bombardment through its imprint on the thermochronologic record of asteroidal meteorites.
We focus on the \kar{} system, which is susceptible to resetting at relatively low temperatures (e.g.~ref.\cite{Bogard1995impact}), and therefore well-suited to capturing the relatively subtle thermal signals of impact-heating~\cite{Davison2012postimpact,Bland2014pressure}.
{We exclude other thermochronometers to focus our study on lower temperature perturbations and limit the dimensionality of our statistical inversion (Methods)}.
Moreover, we follow the logic of ref.\cite{Boehnke2016} that protracted, bombardment-warmed planetary histories systematically young \kar{} system ages.
Although prior evaluation of meteorite thermochronology has constrained dynamical instability to within the first $\sim$100~M\yss{}~\cite{Mojzsis2019onset}, establishing more precise constraints has been challenging because these timescales overlap with those of radiogenic heating and subsequent cooling of the meteorite parent planetesimals.
To deconvolve these endogenic thermal histories of asteroidal meteorite parent bodies from exogenic bombardment reheating histories, we use a Bayesian statistical approach.

We compiled a database of cooling ages from the thermochronologic \kar{} system for chondrite meteorites with provenance in the inner solar system~\cite{Kruijer2020great}: the ordinary chondrites (OC) comprised of the H, L, and LL types; the enstatite chondrites (EC) comrpised of the EH and EL types; and the Rumuruti-type chondrites (RC).
These groups come from undifferentiated planetesimals with well-characterized thermal histories largely explained by simple conductive cooling histories~\cite{Miyamoto1981}, which are relatively straightforward to model (Methods).
The various parent bodies corresponding to these meteorites likely experienced similar thermal histories, and our Bayesian statistical approach allows us to rigorously account for uncertainties stemming from modest differences among parent body histories (Methods). 
While prior work has explored the chronologies of highly shocked and impact-melted chondrites~\cite{Bottke2015}, we consider all chondritic \kar{} system ages, since shock effects and shock heating are heterogenously distributed~\cite{Bland2014pressure,Stoffler1991} and seemingly unshocked meteorites may still record heating from impacts.
This approach allows us to use a database that spans solar system history, with a high density of ages throughout the timescales of potential GPM (Figs.~\ref{fig:timescales}, \ref{fig:dates}).

To resolve the respective contributions of endogenic and exogenic heating on the chondritic \kar{} record, we explore the parameter-space describing this two-component thermal history with a Markov chain Monte Carlo (MCMC) algorithm (Methods). This model is constrained by two sets of priors: our database of \kar{} system ages (Fig.~\ref{fig:dates}) and published constraints on the thermochronologic model parameters (Table~\ref{tab:params}). 
Through its exploration of the parameter space, the algorithm produces Markov chains whose stationary distributions yield Bayesian posterior estimates of each parameter in the simulation.
To constrain the timescales of GPM and dynamical excitation, the simulated thermochronologic histories include exponentially decaying bombardment fluxes, which may mimic either collisional pulses resulting from GPM or a more secular decline in collisions as the solar system ages (Extended Data Fig.~\ref{fig:model}). 
In the following sections, we report and discuss the results of these simulations and their implications for the GPM history of the solar system.

\section{Results}\label{results}
\subsection{Database of \kar{} system ages}

\begin{figure}
    \centering
    \includegraphics[width=\textwidth,keepaspectratio]{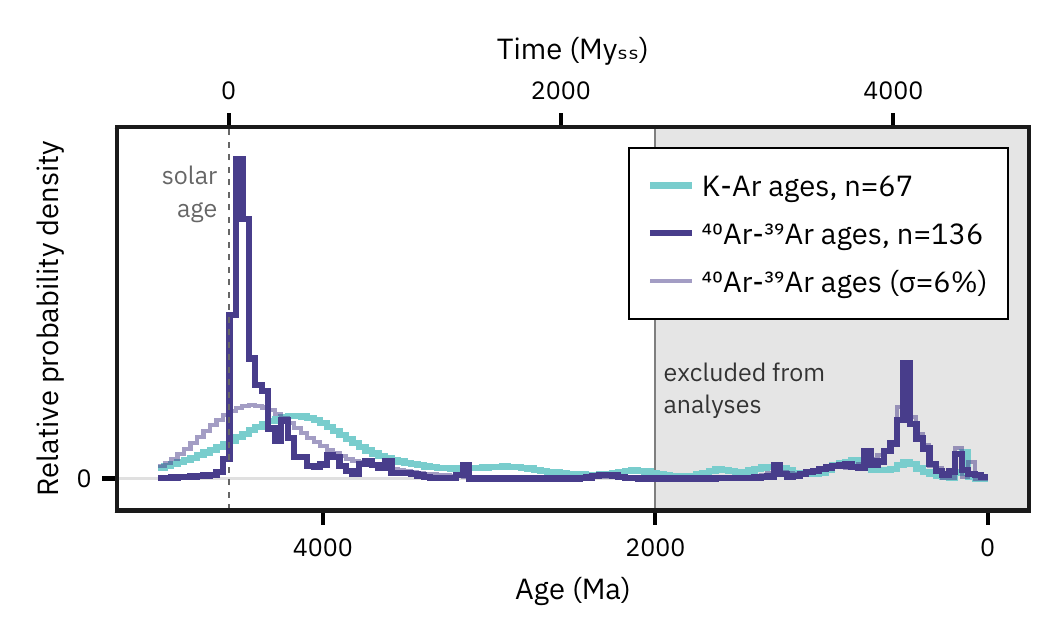}
    \caption{
    \textbf{Distributions of \kar{} system ages, measured by the K-Ar and \arar{} methods.} 
    Time is reported as both age (Ma, lower x-axis) and time after solar system formation (M\yss{}, upper x-axis). 
    Histograms reflect the summed distributions of ages from each dataset. 
    K-Ar age density has a broad, shallow maximum at $\sim$4200~Ma, that gradually decays to low probability densities with minor local maxima extending to the present day. 
    \arar{} age density monotonically decreases from a sharp peak ($>2\times$ the height of the K-Ar peak) shortly after the age of the solar system (dashed line), with minor local peaks between $\sim$4200 and 3600~Ma. 
    By 3500~Ma, the distribution converges on near-nil probability density until an approximately symmetric local maximum within the last 2000~Ma.
    Since these $<$2000~Ma ages (shaded) are not associated with giant planet migration (see text), we exclude them from our analyses.
    If we recalculate the \arar{} age distribution with lower precision ($\sigma=6\%)$ to mimic the K-Ar system, the early peak is more broad and shallow but the maximum remains $>$4300~Ma. 
    }
    \label{fig:dates}
\end{figure}

Figure~\ref{fig:dates} shows the distribution of \kar{} system ages in our database (n=203), which includes measurements by the \arar{} and K-Ar techniques (Methods).
\arar{} ages follow a bimodal distribution.
The maximum of its probability density is slightly younger than the solar age and declines monotonically to minimal density by $\sim$3~Ga, followed by a secondary peak of ages at $\lesssim$1~Ga.
The distinct bimodality of \arar{} ages results from the technique's higher analytical resolution and capability for identifying and excluding disturbed K-bearing domains from age calculations (e.g.~ref.\cite{Turner1978early}; see Methods for a more comprehensive explanation of the two methods).
In contrast, the peak K-Ar age density is $>$300~Ma younger than the \arar{} peak and has a broad noisy tail that extends to the present-day, which reflects the propensity of K-Ar ages for partial resetting by low-temperature Ar-loss (Methods). 
Even when \arar{} ages are resampled at the lower precision of K-Ar ages (on average $\sigma=6\%$), it produces a broader and shallower density profile, but its maximum is still older and more pronounced than the maximum of the K-Ar age distribution~(Fig.~\ref{fig:dates}).
Since K-Ar ages are systematically younger due to partial resetting, we only include \arar{} ages in our analyses and interpretations.

The recent peak in \arar{} cooling ages at 500~Ma is comprised only of the OC groups, represented predominantly by the L chondrites (Extended Data Fig.~\ref{fig:agegrid}). 
This is consistent with the $\sim$460~Ma disruption of an L chondrite parent body inferred by prior studies~\cite{Bogard1995impact,Korochantseva2007lchondrite}.
Since the \kar{} system is resilient to resetting during atmospheric entry~\cite{Turner1990retention,McConville1988laser}, we are confident that these $<$2~Ga cooling ages reflect collisional heating events rather than modern resetting.
Thus, we interpret the $\sim$500~Ma secondary peak of \arar{} cooling ages as a relatively recent collisional event---or events---that comminuted asteroids and left meteoroid rubble piles on near-resonant orbits, which were later perturbed onto Earth crossing trajectories \cite{Morbidelli1998orbital}. 
The apparently coordinated $\sim$500~Ma reheating of OCs implies a relationship in their excavation and delivery to Earth. 
We suggest that this coordination may reflect either an orbital relationship among discreet H/L/LL parent bodies or a shared provenance of these stones from an L-dominant OC rubble pile parent body.
Since these young impact-heating ages occur on timescales substantially post-dating any plausible GPM processes (Fig.~\ref{fig:timescales}), we exclude all $<$2~Ga \arar{} ages from our analyses.
Table~\ref{tab:ages} reports population statistics of the $>$2~Ga \arar{} ages that we use as priors in our Bayesian inversion calculations.

We interpret the ancient peak and decline in \arar{} cooling ages to reflect endogenic thermal histories overlain by exogenic reheating. 
We associate the exogenic reheating with impacts related to secular collisional history and/or early bombardment(s)~\cite{Mojzsis2019onset}. 
An LHB predicts an enhancement in bombardment $\le$4~Ga, which would appear as a distinct peak in cooling ages at its onset, but the predominantly monotonic decline from 4.5--3.0~Ga contradicts such a history.
Minor local maxima in the distribution around 4~Ga may reflect a subtle LHB signal (Fig.~\ref{fig:dates}), which we rigorously test (and reject) with the Bayesian inversion we apply herein.


\subsection{Thermochronologic evidence for a bombardment history}

Thermochronologic simulations of an unperturbed, radiogenically heated body produce a monotonic decline of cooling ages following an initial peak (Extended Data Fig.~\ref{fig:model}). 
To test whether or not these features of the $>$2~Ga \arar{} age distribution require reheating by impacts, we run MCMC inversions with and without any bombardment histories.
We concurrently use two sets of priors: the distribution of measured  $>$2~Ga \arar{} ages and constraints on the thermochronologic model parameters (Table~\ref{tab:params}).
Simulations that are inconsistent with the prior information (i.e. too many/few bombardments) yield posteriors in poor agreement (hereafter ``tension'') with those priors, whereas scenarios that are consistent with prior information converge on posterior distributions that are largely concordant with priors.

\begin{figure}
    \centering
    \includegraphics[width=\textwidth,keepaspectratio]{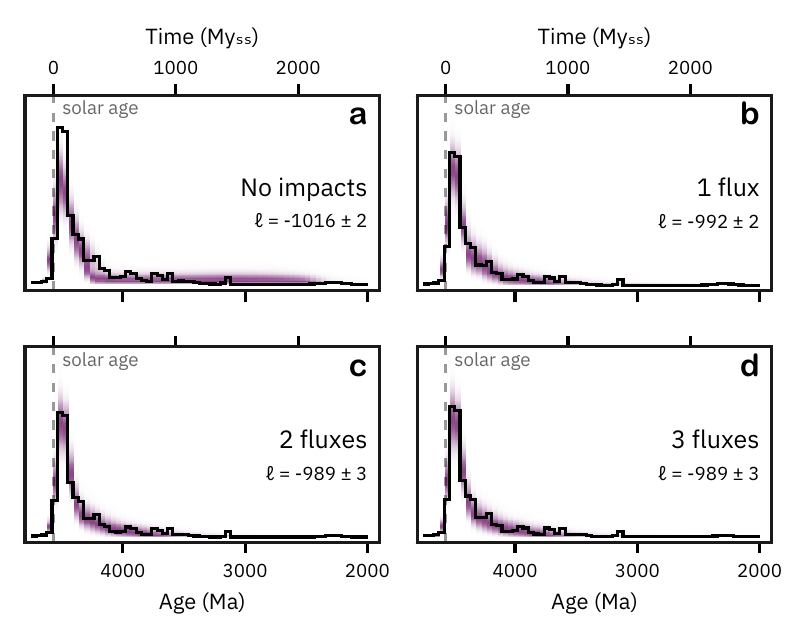}
    \caption{
    \textbf{Comparison of prior (measured, black) and posterior (modeled, purple) \arar{} age distributions for different bombardment histories.}
    Time is reported as both age (Ma, lower x-axes) and time after solar system formation (M\yss{}, upper x-axes). 
    The dashed lines demarcate 0~M\yss{}.
    The purple heatmaps show the relative density of posterior age distributions (darker = higher density), overlain by the prior distribution of measured ages in black (as in Fig.~\ref{fig:dates}). 
    (\textbf{a}) A no-impact scenario  yields a relatively poor agreement between measured and simulated thermochronologic ages, corresponding to a mean log-likelihood ($\ell$) of $-1016 \pm 2~(1\sigma)$. 
    (\textbf{b}--\textbf{d}) Scenarios with impact fluxes yield posterior distributions that are concordant with the prior, corresponding to $\ell=-992 \pm 2$ for a single impact flux scenario (\textbf{b}) and $\ell = -989 \pm 3$ for scenarios with 2 or 3 impact fluxes (\textbf{c},~\textbf{d}).
    To mimic the contribution of uncertainties in the prior distribution, each posterior age distribution (n=$10^6$) is recalculated by randomly resampling 100 draws with the mean uncertainty of the measured \arar{} dates ($\sigma\sim1\%$) and calculating a histogram (as in Fig.~\ref{fig:dates}) from the resampled ages. 
    }
    \label{fig:fits}
\end{figure}

Figure~\ref{fig:fits} compares the prior distribution (measured) with the posterior (simulated) distributions of \arar{} ages from simulations with 0--3 bombardment episodes. 
Simulations without impact reheating (Fig.~\ref{fig:fits}a) yield posterior distributions of \arar{} ages that fail to reproduce the brief early peak of ages, the shape of the monotonic decline, and the near-nil density of ages between 3.5--2~Ga.
The posterior distributions of several model parameters are in stark tension with their priors (Extended Data Fig.~\ref{fig:0Fposts}, Extended Data Table~\ref{tab:posts01}).  
In contrast, each of the simulated asteroid thermal histories that incorporate impact reheating (Fig.~\ref{fig:fits}b--d) yield posterior distributions that are concordant with the prior distributions of \arar{}-ages and model parameters.
These observations are quantitatively supported by the corresponding log-likelihoods ($\ell$), such that a no-impact history corresponds to a mean $\ell = -1016 \pm 2~(1\sigma)$, whereas bombardment histories yielded similar $\ell > - 1000$.
We conclude that the meteorite record requires one or more $>$2~Ga bombardment events to explain the chondritic \arar{} age distribution.

\subsection{Impact flux parameters and their characteristics}

We simulate episodes of enhanced collisions and bombardment in the asteroid belt with exponentially decaying fluxes characterized by an onset date ($t_o$), initial impactor flux ($F_o$), and $e$-folding timescale ($\tau$).
We simulate impact reheating nonphysically (Methods), so the values of $F_o$ are proportional to an impact flux rate and reflect the degree of thermochronologic resetting but ought not be interpreted as quantitative estimates of asteroid belt impact flux.
Our model framework accommodates up to three bombardment events, which we denote with $\alpha$, $\beta$, and $\gamma$.
To limit multimodality in the posterior distributions of multi-bombardment simulations, we require the onset dates of these bombardments to occur sequentially ($t_o\alpha<t_o\beta<t_o\gamma$). 
Additionally, in simulations with $>1$ bombardment we assign the first bombardment ($\alpha$) as a primordial bombardment ($t_o=0$~M\yss{}) to reduce dimensionality. 
We envision this primordial flux as the long-term (Gy-scale) background rate of inner solar system collisions (as in refs.\cite{Wetherill1977evolution,Nesvorny2017modeling,Brasser2020impact}), distinct from punctuated and transient GPM-triggered bombardment.
For this reason, we require that the $\tau$ of the primordial flux exceeds those of the post-accretion bombardments meant to simulate the dynamical consequences of GPM.

In all simulations, we find that $F_o$ and $\tau$ are closely related.
Within any given bombardment event, $F_o$ decreases with increasing $\tau$ (Fig.~\ref{fig:impacts}, Extended Data Figs.~\ref{fig:1Fcorner}--\ref{fig:3Fcorner}).
This inverse scaling of $F_o$ and $\tau$ indicates that a longer bombardment duration can compensate for a relatively low initial flux while, conversely, rapid recovery can compensate for a high initial flux. 
Among bombardment events, each bombardment event falls into one of two categories with respect to $F_o$ and $\tau$: intense and brief bombardments (hereafter intense/brief) are characterized by $F_o > 100$~My$^{-1}$ and $\tau \ll 100$~My (e.g. Fig.~\ref{fig:impacts}a), whereas mild and protracted bombardments (hereafter mild/protracted) are characterized by $F_o \ll 100$~My$^{-1}$ and $\tau > 100$~My (e.g. Fig.~\ref{fig:impacts}b).
Additionally, $t_o$ scales inversely with $F_o$, such that more intense bombardment events occur earlier in solar system history. 
Notably, parameters from separate bombardment events never appear correlated, indicating that they are independent with respect to each other.
These patterns are consistent among all simulations that include bombardment events. 

\subsection{Number, timescale, and intensity of bombardment events}

A single-bombardment scenario converges on posterior impact flux parameter distributions that reflect a mild/protracted bombardment (Extended Data Fig.~\ref{fig:1Fcorner}), such as that expected for the background rate of inner solar system collisions. 
The persistence of the mild/protracted impact flux over the intense/brief flux emphasizes the critical importance of secular collisions to explain the thermochronologic histories of chondrites as well as the inner solar system impact record~\cite{Nesvorny2017modeling,Brasser2020impact}.

Numerous lines of independent evidence indicate that there was an episode of GPM in solar system history that would have driven extensive small-body migration through the region of the asteroid belt \cite{DeMeo2014solar,Kruijer2017age,Gomes2003origin,Tsiganis2005,Gomes2005origin,Clement2018mars,Clement2019excitation} and thereby precipitated a punctuated episode of asteroid collisions. 
Therefore, a single-bombardment scenario is insufficient to meet the requisite dynamical history of the solar system, and at least one more impact flux is required.
In support of this conclusion, the posterior distributions of several model parameters are in greater tension with their priors for single-bombardment simulations than those with 2--3 bombardment histories (Extended Data Figs.~\ref{fig:1Fposts}, \ref{fig:2Fposts}).

Figure~\ref{fig:impacts} reports the posterior distributions of bombardment parameters for simulations characterized by two bombardments: a primordial bombardment anchored to the solar age (0~M\yss{}) and a post-accretion bombardment with an onset date ($t_o$) that varies along with all other parameters in Table~\ref{tab:params}.
The post-accretion bombardment (Fig.~\ref{fig:impacts}a) is intense/brief with a median onset date of 
{$11.3^{+9.5}_{-6.6}$~M\yss{} (50\% credible interval, hereafter CI; $^{+15.0}_{-8.6}$~M\yss{} at 68\% CI, $^{+44.4}_{-11.0}$~M\yss{} at 95\% CI)} and mean of $15\pm14$~M\yss{} ($1\sigma$).
We favor the median as an estimate of central tendency, since the distribution is skewed.
In contrast, the primordial flux (Fig.~\ref{fig:impacts}b) is mild/protracted. 
These results are consistent with a solar system history characterized by a slowly decaying background rate of infrequent collisions punctuated by an early ($<100$~M\yss{}) episode of dynamical excitation and intense bombardment \cite{Mojzsis2019onset,Nesvorny2017modeling}.
Moreover, this intense and transient collisional episode satisfies theoretical predictions of GPM-associated bombardment in the inner solar system~\cite{Gomes2005origin,Walsh2011,Clement2019excitation}.

We also explore the effect of incorporating a third impact flux $\gamma$ (also post-accretion) and find that three bombardment events reproduce the \arar{} age distribution as well as two (Fig.~\ref{fig:fits}). 
In this scenario, the  $\alpha$ (primordial) and $\beta$ (post-accretion) impact fluxes are nearly identical to the respective fluxes of the two-bombardment simulation shown in Fig.~\ref{fig:impacts} (Extended Data Fig.~\ref{fig:3Fcorner}). 
The posterior of the $\gamma$ flux is bimodally distributed with respect to $t_o$ and $\tau$. 
One mode mimics the $\beta$ bombardment with a local maximum at $t_o \sim 10$~M\yss{} and $\tau \sim 10$~My.
The other mode is extremely late in the simulation time-domain with $t_o > 1000$~M\yss{}---perhaps capturing one of the minor maxima between 3.5 and 2~Ga---and has an extremely brief lifetime of $\tau \sim 0.01$~My (i.e. flux becomes negligible within a single 1~My timestep).
Thus, the third flux either mimics the $\beta$ event or is suppressed by an extremely short $\tau$.
Since a third flux neither adds new information nor has any discernible effect on the simulated age distribution (Fig.~\ref{fig:impacts}), we conclude that it is unnecessary and do not consider it further.

In our multi-bombardment simulations, the $\alpha$ bombardment $t_o$ is always primordial ($t_o=0$~M\yss{}).
If we instead simulate two post-accretion impact fluxes ($\beta$, $\gamma$) with no primordial bombardment requirement (Extended Data Fig.~\ref{fig:2Fcorner}, Supplementary Fig.~1), we get a similar result to the one-primordial/one-post-accretion two-bombardment scenario in Fig.~\ref{fig:impacts}.
The posterior distributions of $\beta$-flux parameters are indistinguishable between these scenarios, and the distributions of $F_o$ and $\tau$ for the $\gamma/\alpha$-fluxes are very similar.
Curiously, in scenarios with only post-accretion bombardments, the $t_o$ of the mild/protracted fluxes have median values between 100--140~M\yss{}, rather than a more ``primordial'' value near 0~M\yss{}.
In the case of the 2-bombardment distribution, the late onset date partly reflects the model requirement that $t_o\beta < t_o\gamma$.
Since there is no such restriction in the case of the 1-bombardment distribution, the inversion likely favors the later onset because it produces the long, shallow tail of cooling ages to 3.5~Ga without also placing a larger, long-lasting impact flux overlapping with the timeframe of maximum age density at $\sim$4500~Ma (Fig.~\ref{fig:dates}). 
Nonetheless, whether the mild bombardment is primordial or post-accretion does not discernibly affect the characteristics of intense bombardment or the concordance between the observed and simulated \arar{} age distributions ($\ell=-990 \pm 3$ for two post-accretion bombardments).
Thus, our assumption of a primordial $\alpha$ bombardment is valid for the purposes of this study, and we favor this approach since it is more consistent with the mild/protracted impact flux representing a secular process.

\subsection{Comparison to non-bombardment parameters}

For all bombardment scenarios, posterior distributions of environmental, cosmochemical, asteroidal, and material parameters are overall concordant with their prior constraints (Extended Data Fig.~\ref{fig:2Fposts}, Extended Data Table~\ref{tab:posts23}, Supplementary Figs.~1--2). 
Our models' ability to satisfy both these independent constraints as well as the chondritic \arar{} record supports their overall veracity.
We briefly consider three minor discrepancies between the priors and posteriors.
First, accretion time ($t_a$) tends toward slightly later values, likely reflecting a bias towards the later accretion timescales for the parent bodies of OCs~\cite{Sugiura2014correlated}, which comprise $>$70~\% of the \arar{} database. 
Second, the disk midplane temperature ($T_m$) tends toward lower values than the observation-constrained prior~\cite{Woolum1999astronomical} but remains consistently within chondritic constraints of $<$503~K~\cite{Schrader2018background}.
Since the relatively warm gaseous disk dissipated early on in the asteroids' thermal histories, the tendency toward lower temperature may reflect the unimodal model compensating for this bimodal temperature history, though insulation from surface regolith layers likely reduced the effect of the temperature drop.
Third, effective Ar closure temperatures ($T_c$) trend very slightly toward higher values compared to the expected distribution constrained from Ar degassing data~(Table~\ref{tab:params}), suggesting that effective cooling rates were slightly faster than those assumed in prior studies (e.g.~ref.\cite{Turner1978early}).

\section{Discussion}\label{disc}
\begin{figure}
    \centering
    \includegraphics[width=.75\textwidth,keepaspectratio]{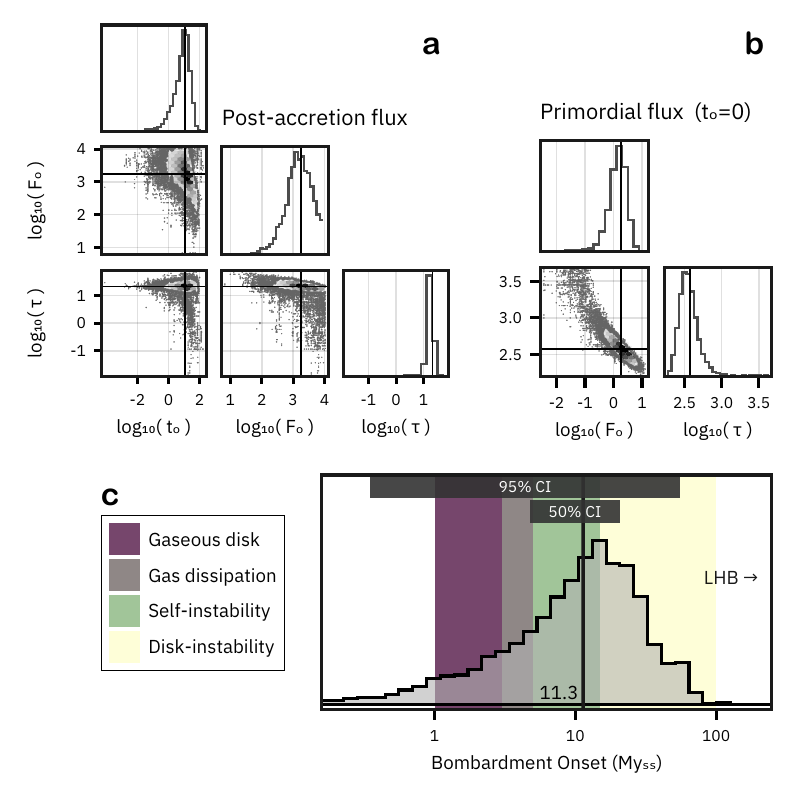}
    \caption{
    \textbf{Posterior distributions of simulated bombardment history parameters.}
    Panels \textbf{a} and \textbf{b} are corner plot diagrams: diagonals depict one-dimensional histograms of each parameter, and off-diagonals depict the 2-dimensional distributions of each parameter pair, with heatmaps of distribution density within 2$\sigma$ of the means (darker cells reflect higher density).
    Black lines trace the median values of each parameter.
    (\textbf{a}) corresponds to a ``post-accretion'' flux, for which the Bayesian inversion explores its onset time $t_o$ (M\yss{} = My after CAIs), initial impact flux $F_o$ (My$^{-1}$), and $e$-folding time $\tau$ (My).
    $F_o$ tends to decrease as $\tau$ and $t_o$ increase.
    (\textbf{b}) corresponds to a ``primordial'' flux where the inversion explores values of $F_o$ and $\tau$, but $t_o$ (not shown) is anchored to 0~M\yss{}. 
    $F_o$ and $\tau$ exhibit a pronounced inverse relationship.
    Notably, the median values of $F_o$ and $\tau$ for the post-accretion flux (\textbf{a}) are respectively 1000$\times$ and $<$0.1$\times$ the primordial flux (\textbf{b}).
    Summary statistics are tabulated in Extended Data Table~\ref{tab:posts23}.
    (\textbf{c}) Comparison of the posterior distribution of bombardment onset date ($t_o$ in \textbf{a}) with the timescales of potential giant planet migration stimuli, as in Fig.~\ref{fig:timescales}.
    Dark bars demarcate the 95~\% and 50~\% credible intervals (CI). 
    The lower 26~\% of the distribution overlaps the timescales of gas-disk-driven migration, the uppermost 38~\% overlaps the timescales of planetesimal disk-triggered instabilities, and 45~\% of the distribution interior---including the median (11.3~M\yss{}, vertical line) and mean (15.0~M\yss{})---overlaps the timescales of gas dissipation and self-triggered instabilities.
    The hypothesized Late Heavy Bombardment (LHB) timescale lies beyond the domain.
    Distributions reflect $10^6$ Markov chain steps. 
    }
    \label{fig:impacts}
\end{figure}

Since a two-bombardment history is necessary and sufficient to explain the chondrite thermochronologic record in the context of post-accretion dynamical instability (Results), we select the scenario shown in Figs.~\ref{fig:fits}c and \ref{fig:impacts} (two bombardments: one primordial, one post-accretion) as our preferred model to interpret the timescales of giant planet migration (GPM).
{The posterior $e$-folding timescales of the mild/protracted primordial fluxes modeled here ($\sim$380~My) are consistent with those of the long-term ``tails'' predicted for the inner solar system cratering record (e.g.~200--400~My)~\cite{Brasser2020impact} due to the gradual dynamical ``leaking'' of asteroids on Mars-crossing orbits onto planet-crossing trajectories \cite{Wetherill1977evolution}.}
Since this primordial impact flux reflects a secular rate of collisions over solar system history, we conclude that it is not related to GPM.
In contrast, the intense/brief impact flux is consistent with a violent yet transient bombardment event. 
Though our parameterization uses arbitrary units, the median $F_o$ of this bombardment heats $\sim$50\% of the simulated asteroid volume at the onset of bombardment.
Although the upper constraints on $F_o$ are poorly constrained by our model and may not be representative of the body's deeper interior (see extended discussion in Methods), the large values underscore the severity of this event, especially for shallower portions of early chondritic bodies.
Additionally, this bombardment dissipates on a timescale similar to the most intense and short-lived component of impact flux models for the inner solar system~\cite{Nesvorny2017modeling}, which are comparable to the timescales predicted for dynamical cooling after GPM~\cite{Gomes2005origin,Walsh2011}.
This concordance motivates our first of two key conclusions: the thermochronologic record of the asteroid belt records a single GPM event.
The second key conclusion regards its timing.

Figure~\ref{fig:impacts}c compares the posterior distribution of the GPM-induced bombardment onset to the timescales of potential stimuli, as described above and in Fig.~\ref{fig:timescales}.
The timescales of GPM pre-date canonical LHB timescales by hundreds of My (Figs.~\ref{fig:timescales}, \ref{fig:impacts}).
Since bombardment of the terrestrial planets by asteroids or comets requires scattering within or through the asteroid belt, we expect that at least one of the $\sim$6 parent bodies represented in our database would record these collisions.
Our findings firmly refute any intense bombardment occurring $\gg$100~M\yss{}, consistent with other recent work~\cite{Mojzsis2019onset,Nesvorny2018evidence,RibeirodeSousa2020,Fassett2012lunar,Boehnke2016,Xie2023new}.

We do not resolve a {significantly} unique timeframe for $<$100~M\yss{} GPM, since all 4 timescales fall within the traditional 95\% CI. 
However, only the lower 26\% of the distribution overlaps timeframes during which the gaseous disk could have played a causal role in GPM ($<$5~M\yss{}, Fig.~\ref{fig:impacts}). 
Observations of protoplanetary disks for solar-type stars show that disk lifetimes are consistently $<$10~My, with few disks surviving $>$6~Ma and most dissipated within $\lesssim$3~Ma \cite{Haisch2001disk,Sung2009spitzer,Williams2011protoplanetary}.
Dynamical models corroborate these observations and predict mean disk lifetimes of 3.7~Ma \cite{Li2016lifetimes,Kimura2016birth}. This is consistent with meteoritical evidence for gas disk dissipation beyond Jupiter's current orbit ($\sim5$~au) by 3.5--5~M\yss{}~\cite{Borlina2022lifetime}. 
Thus, we approximate the plausible timescales of gas dissipation to 3--5~M\yss{}. Since a Grand Tack-style migration must be shortly followed by gas dissipation to halt the outward migration of Jupiter and Saturn, we assign a generous lower-limit for gas-embedded GPM onset at 1~M\yss{} (Fig.~\ref{fig:impacts}c).
This 1--5~M\yss{} timeframe for a Grand Tack style migration overlaps with only 20\% of the posterior distribution of onset times, and even this is an upper limit due to potential insensitivity of the Bayesian inversion at very early timescales (see extended discussions in Methods).
{We conclude that a Grand Tack style migration was unlikely.}

The remaining upper 75\% of the distribution overlaps timescales post-dating dissipation of the gaseous disk (Fig.~\ref{fig:impacts}c). 
The 50\% CI of the distribution (4.75--20.76~M\yss{}), including the mean (15.0~M\yss{}) and median (11.3~M\yss{}), overlaps the timescales of instability resulting from self-unstable orbital architectures left behind after dissipation of the gaseous disk. 
{This interval is consistent with Pd-Ag cooling ages of iron meteorites ranging from $\sim$7.8--12.8~M\yss{} that reflect widespread energetic asteroid collisions in this timeframe~\cite{Hunt2022dissipation}.}
The upper 38\% of the posterior overlaps the timescales of GPM caused by interaction with an outer planetesimal disk~\cite{RibeirodeSousa2020}.
{Collectively, $>$75\% of the distribution overlaps timescales associated with dynamical instability of giant planets ($>$3--5~M\yss{}) rather than a Type~II style migration embedded in a gaseous disk ($\le$5~M\yss{}).}
This motivates our second key conclusion that GPM in our solar system was {likely} a result of a dynamical instability occurring shortly after the dissipation of the gaseous disk.
Though the results are non-unique {at the canonical 95\% CI}, the most probable mechanistic stimulus of instability was an unstable giant planet orbital configuration, potentially exacerbated by interaction with a massive outer disk of planetesimals~\cite{RibeirodeSousa2020}.
Whether the systems were destabilized by asymmetric torques of a receding disk~\cite{Liu2022early}, inherently unstable orbital configurations without the support of a gaseous disk~\cite{RibeirodeSousa2020}, interactions with an outer planetesimal disk~\cite{Tsiganis2005,Clement2019excitation,RibeirodeSousa2020}, or any combination of these processes,  gas dissipation was a critical process in triggering or predisposing the system to GPM.

Our results provide a cosmochronological constraint on the timescales of GPM in our solar system, constrained with quantitative precision by a broad subset of the meteorite record (n=97).  
We expect that our results will be refined as future efforts expand meteorite thermochronometry, provide greater astronomical context for this data (e.g. asteroid sample return missions), and improve thermal models to simulate more physical scenarios. 
{In particular, improved \kar{} decay constant calibration at early solar timescales and more nuanced understanding of bulk chondritic Ar release patterns will pave the way for substantially improving the quality of our database of ages and inversion results (Methods).}
We recognize that the model is a simplification of meteorite thermal histories. 
Its greatest limitation is its relative insensitivity  to impact heating prior to thermochronologic closure (see Methods for comprehensive treatments of all assumptions and limitations).
Although over-early ($<$1~M\yss{}) bombardment scenarios are already incompatible with plausible solar system GPM chronologies (Fig.~\ref{fig:impacts}), we expect that simulations incorporating more nuanced impact processes and numerically solved thermal histories may solidify and refine our present findings. 
Yet, despite its simplicity, the model effectively replicates the broad \arar{} thermochronologic record of chondrites and corroborates extant constraints on chondritic planetesimals (Table~\ref{tab:params}), both of which support its overall veracity for early solar system chronology.

GPMs appear to be a quotidian feature of exoplanetary systems \cite{Pu2015,Raymond2010planetplanet}. Our findings further solidify the growing evidence that this process is characteristically constrained to the earliest stages of a planetary system's history.
We encourage future dynamical studies to focus on this epoch of planetary evolution, as it likely plays an outsized role on the long-term physical and chemical structure of exoplanetary systems, as it did in ours.
Similarly, our results motivate observational emphases on young ($<$20~My) exoplanetary systems (e.g.~refs.\cite{Rizzuto2020tess,Mann2022tess}) that might host planetary architectures on the verge of or recovering from instability and migration.

\section{Methods}\label{methods}
\subsection{Database of \kar{} system ages}
To evaluate the thermal histories of undifferentiated bodies in the solar system, we compiled a database of \kar{} system cooling ages for chondrites with inner solar system provenance (ordinary, enstatite, and Rumuruti-type).
\K{40} undergoes branching decay to \Ar{40} and \iso{40}{Ca} with a half-life of $\sim$1.25~Gy.
The \kar{} decay system is a thermochronometer due to the temperature dependent diffusivity of gaseous Ar through mineral crystal lattices. 
The nominal closure temperature of plagioclase (the most K-rich mineral found in chondrites) to Ar diffusion is approximately 500~K for cooling rates spanning 1--1000~K/My \cite{Turner1978early,Bogard2009arar}.

Dates in the \kar{} system have been measured by one of two techniques: the K-Ar and \arar{} methods.
The K-Ar method entails first degassing Ar and measuring its isotopic composition to obtain the total abundance of \Ar{40}, then measuring the K abundance of the degassed sample and correcting for a known or assumed \K{40}/K to obtain the absolute abundance of \K{40}, and finally calculating a date from the ratio \Ar{40}/\K{40}.
While the earliest \kar{} system ages were measured by this method, K-Ar ages are inaccurate and misleading if the system was previously heated and partially degassed, which results in ages that fall between the primary cooling age and the age of reheating. 
To circumvent this issue, the \arar{} method entails irradiating a sample with fast neutrons to convert \K{39} to \Ar{39} and then degassing the sample and measuring the isotopes of Ar.
Since \K{40}/\K{39} only varies substantively due to time-dependent \K{40} decay,  \K{39} (and therefore irradiated \Ar{39}) is a reliable proxy for \K{40} for a given moment in geologic time.
A standard of known age is irradiated alongside the unknown sample and its measured \Ar{40}/\Ar{39} is used to correct for the incomplete conversion of \K{39} to \Ar{39}.
Thus, paired measurements of \Ar{40}/\Ar{39} for a sample of unknown age and a standard may be used to calculate the unknown sample's \Ar{40}/\K{40} and age. 
By sequentially degassing samples at a range of temperatures with the \arar{} method, phases that experienced partial loss of Ar at lower temperatures can be identified and excluded from the final age calculation. 

We compiled a database of both K-Ar and \arar{} cooling ages from the published literature, using several prior compilations as a starting point~\cite{Bogard1995impact,Swindle201440ar,Mojzsis2019onset}.
In most cases, we followed the recommended ages reported by the publishing authors, typically plateau or ``reduced plateau'' ages in the case of the \arar{} method.
If there were two clear \arar{} plateaus, we incorporated the older, as this study focuses on the early cooling history of asteroids and the younger cooling ages (typically $<1$~Ga) typically reflect collisions that ejected meteoroids from larger parent bodies toward Earth-crossing orbits~\cite{Bogard199539ar40ara}.
However, the uncertainties of dates were not always reported or calculated from rigorously propagated uncertainties.
We do not include ages that are reported as only minimum/maximum ages, as quantified uncertainties are necessary for our Bayesian approach. 
Where dates were given without uncertainty or calculated from an assumed K abundance, we assumed a cautiously large uncertainty of $2\sigma = 10\%$.
In cases where multiple ages were reported and there was not a clearly more contemporary or less disrupted age (n=13), we calculated the mean of the reported ages by Monte Carlo method.

\subsubsection{Using common constants and re-calibrating \kar{} ages}

For quantitative comparison, all the ages in our database must be calculated relative to a common set of decay constants and \K{40}/K. 
The branching decays of \K{40} to \Ar{40} by electron capture and \iso{40}{Ca} by $\beta^-$-emission are respectively described by the decay constants $\lambda_e$ and $\lambda_\beta$ (the summed decay constant is denoted $\lambda$).
However, the values of these constants used to calculate \kar{} ages have changed over the history of the system's geochronometric use.
The decay constants of ref.\cite{Steiger1977subcommission}~(hereafter SJ77) have been used almost ubiquitously by the meteoritics community since its publication, despite publication of more recently revised \K{40} decay constants \cite{Renne2010joint,Renne2011response}.
This is in part due to uncertainty regarding the accuracy of \K{40} decay constants on early solar system timescales {(the decay constants of ref.\cite{Renne2010joint,Renne2011response} were calibrated to $<$3~Ga standards) and debate over the appropriate approach to recalibration for timescales $\ge$4~Ga}~\cite{Schwarz2011comment,Renne2011response}.
However, a recently reported \arar{} age calculated with the SJ77 decay constants for a ureilitic clast (MS-MU-011) of the Almahata Sitta meteorite agrees with its corresponding Pb-Pb age nearly within $1\sigma$~\cite{Turrin202340ar}.
Since this fast-quenched system likely cooled through effective Pb and Ar closure almost simultaneously~\cite{Turrin202340ar}, the SJ77 decay constant seems more accurate on early solar system timescales than previously argued.
Additionally, the ubiquitous use of SJ77 combined with inconsistent reporting of co-irradiated standard information across much of the literature inhibits accurate recalibration.
Thus, we conclude that the SJ77 decay constants are the best option available at this time for calculating \kar{} system ages of meteorites, and note that errors stemming from using the different decay constants ({$<$22}~Ma) are well within the 1-$\sigma$ uncertainties of most ages in our database.
{Nonetheless, we explore the effects of using alternative calibration schemes below.}

To ensure that all \kar{} ages used in this study are calculated relative to a set of common decay constants and \K{40}/K ratio ($K$), we recalculate all \arar{} and K-Ar ages published prior to 1977 with the values of SJ77. 
Given the two different methods employed and non-standardized reporting of methods, these recalculations required one of three scenarios, each of which we employed using a Monte Carlo method to propagate all reported uncertainties.

The first and most straightforward scenario entails recalculating K-Ar ages.
To do this, we rearrange the \kar{} age equation to calculate the measured \Ar{40}/\K{40} ratio from the reported age, correct for the updated $K$ ratio, and recalculate the age:

\begin{equation}\label{eq:kar-recal}
    t = 
    \frac{1}{\lambda} \left(
        ln\left[ 
            \frac{\lambda}{\lambda_e}~
            \frac{K'}{K}~
            \frac{\lambda_e'}{\lambda'}~
            (e^{t' \lambda'} - 1)
        \right] + 1
        \right)
\end{equation}
where the prime symbol indicates the reported age and its corresponding decay and $K$ constants, which were consistently (within rounding error) the values used by \cite{Husain197440ar39ar}.

In the case of the \arar{} method, dates are calculated relative to a co-irradiated standard of known age by the equation:
\begin{equation}\label{eq:arar-age}
    t=  \frac{1}{\lambda} 
        ln\left[
            1 + J ~ \frac{^{40}Ar}{^{39}Ar}
        \right]
\end{equation}
where $J$ is calculated from the standard's age, $t_s$, and its measured \Ar{40}/\Ar{39} ratio:
\begin{equation}\label{eq:J}
    J= \frac{^{40}Ar}{^{39}Ar}~(e^{\lambda t_s} -1 )
\end{equation}

Thus, an \arar{} age may be recalculated simply by correcting the J term:
\begin{equation}\label{eq:arar-recal}
    t = \frac{1}{\lambda}~ln[1 +  k_J~(e^{\lambda' t'}-1)]
\end{equation}
where $k_J$ is the ratio of the recalculated and original $J$-factors ($J/J'$). So, the second and third scenarios depend on how the standards used to calculate the $J$ term are calibrated and how methods were reported in each corresponding study.

The second scenario entails cases where $t_s$ was calculated with a chronometric system other than the \kar{} system. Since $t_s$ is constrained independent of the \K{40} decay constants used, $k_J$ is simply calculated by
\begin{equation}\label{eq:kj-indage}
    k_J = \frac{e^{\lambda t_s}-1}{e^{\lambda' t_s}-1}
\end{equation}

The third scenario entails cases where $t_s$ was calibrated by the K-Ar method, as was typical of \arar{} dates measured early in the history of this technique.
In these situations, we recalculated $t_s$ from the previously used $t_s'$ with Equation~\ref{eq:kar-recal}, and calculated $k_J$ through a slightly modified version of Equation~\ref{eq:kj-indage} (note the $t_s'$):
\begin{equation}\label{eq:kj-ararage}
    k_J = \frac{e^{\lambda t_s}-1}{e^{\lambda't_s'}-1}
\end{equation}
Unfortunately, not all studies report the standard or its corresponding $t_s$.
In these ambiguous cases, we calculated a distribution of $k_J$ values by Monte Carlo method using Equation~\ref{eq:kj-indage}, where $t_s'$ is uniformly distributed over the interval $[0,3)$~Ga, which generously encompasses the full range of standards commonly used for the \arar{} method.
This approach results in a range of values for $k_j$ over the interval $1.04<k_j<1.10$, which results in a minor increase in the uncertainty of the recalculated age. 
For example, for an age of $4500~\pm~20$~Ma, the uncertainty increases to $\sim30$~Ma.
Since most \arar{} dates measured prior to 1977 have analytical uncertainties that are significantly larger than this (typically $>1\%$), the effect was minor.

The complete database and these calculations are available in the \ImpactChron{} package at \url{https://github.com/grahamedwards/ImpactChron.jl/tree/main/data}. 
Supplementary Table 1 is a human-readable table of the \arar{} ages used in this study.

\subsection{Software \& Code}

We invert \kar{} system ages for asteroid formation and bombardment histories using an asteroid-scale thermochronologic simulation coupled to a Markov chain Monte Carlo (MCMC) inversion, written in the Julia Language~\cite{Bezanson2017julia} and contained in the package \ImpactChron{} (\url{https://github.com/grahamedwards/ImpactChron.jl}).
We explain this code in detail in the following sections.
We prepared figures and diagrams with \texttt{Makie.jl}~\cite{Danisch2021makie}, \texttt{Pairplots.jl} (\url{https://github.com/sefffal/PairPlots.jl}),  and Inkscape vector graphics editor.

\subsection{Thermochronologic model}
We simulate the thermal history of a probabilistic asteroid characterized by the thermal histories of the inner solar system chondrites---ordinary (O), enstatite (E), and Rumuruti-type (R). 
This assumption is reasonable given the similarities in the inferred parent body histories of OCs and ECs~\cite{Edwards2020accretion,Blackburn2017accretion,Trieloff2022evolution,Gail2019thermal} as well as the observation that intergroup variability of material properties is typically less than intragroup variability for these chondrite groups~\cite{Macke2010survey,Flynn2018physical,Wach2013specific,Yomogida1983physical,Opeil2010thermal,Opeil2012stony}.
Although the parent body history of the R chondrites is less well-constrained, they represent only a minor portion (n=2) of the \arar{} age database (n=136). 
{We exclude carbonaceous chondrites---which have lower temperature aqueous alteration histories---and achondrites---which underwent partial to complete differentiation---since including these samples would require simulating the thermal histories of several distinct parent bodies (and accompanying parameters) that are poorly constrained compared to the OC, EC, and RC parent bodies. 
Hence, we exclude these groups to prevent an expansion of dimensionality that might leave our inversion (see below) underconstrained, for a relatively minor increase in sample size from the comparatively rare carbonaceous chondrite and achondrite meteorite groups.}

For computational efficiency, we separately model (1.) radiogenic heating and conductive cooling through thermochronologic closure for a body unperturbed by impacts and (2.) impact reheating and resetting of \arar{} ages.
To model the unperturbed thermal histories that would result from radiogenic heating and conductive cooling in the absence of any bombardment, we use an analytical solution to the heat equation for a spherical body with an exponentially decaying heat source~\cite{Hevey2006model,Carslaw1959conduction}. 
The parameterization assumes constant bulk density ($\rho$), specific heat capacity $C_p$, thermal conductivity $K$, ambient temperature $T_m$ (i.e.~the solar system midplane temperature at 2.5~au), and body radius $R$ (Table~\ref{tab:params}). 
Heat production is a function of the time of accretion $t_a$ relative to the solar system age $t_{ss}$, the initial \iso{26}{Al}/\iso{27}{Al} of the solar system $^{26}Al_o$, and chondritic Al abundance $[Al]$ (Table~\ref{tab:params}).
We calculate the time-temperature histories of unperturbed cooling in equally spaced concentric shells of the body, and identify the cooling age of each shell as the first timestep with a temperature below the Ar closure temperature (Extended Data Fig.~\ref{fig:model}a,b).
We calculate the volumetric proportion of each shell to produce a distribution of the volumetric abundance of these cooling ages (Extended Data Fig.~\ref{fig:model}c, black curve).

{To account for sampling biases in our database, we consider both (1.)~the delivery and preservation of asteroidal material to Earth as well as (2.)~the nonrandom sampling of meteorites for isotopic analyses.
We assume that meteoroid sampling and delivery to Earth is pseudorandom and therefore meteorite collections approximately reflect the proportions of proto-asteroidal parent body material. 
In contrast, the selection of meteorites for isotopic analysis and thermochronology are biased and nonrandom. 
For instance, researchers have specifically sought out highly shocked L~chondrites to test the hypothesis of a ca.~500~Ma disruption event~\cite{Bogard1995impact}, which we partly account for by excluding ages $<$2000~Ma.
To more thoroughly account for human-induced bias, we weight our simulated abundances of asteroidal material relative to representation in the \kar{} database.}

While we cannot comprehensively control for heterogenous sampling of asteroidal parent bodies, delivery of meteoroids to Earth, and selection of meteorites for study, we account for heterogenous representation of parent body material in our database by weighting our results by petrologic type, since petrologic type is predominantly controlled by a meteorite's provenance within its parent body (e.g.~ref.\cite{Miyamoto1981}).  
For each shell, we assign a petrologic type based on its peak temperature ($T_{type}$ in $^{\circ}$C): $T_3 \le 600 < T_4 \le 700 < T_5 \le 800 < T_6$, following the recommendations of ref.\cite{Lodders1998planetary} for the maximum temperatures of types~3--4 and a minimum type~6 temperature derived from extensive measurements of type~6 OCs\cite{Slater-Reynolds2005peak}.
Since type~7 and impact-melted chondrites reflect exogenous heating events that are not directly related to provenance within a parent body\cite{Edwards2020accretion}, we do not incorporate these classifications into our weighting calculations.
We weight the calculated volumetric abundances of each volumetric shell in our simulation so that the relative abundances of each assigned petrologic type (the sum of the volumetric abundances of all shells of that type) equals the relative abundance of that petrologic type in the \arar{} age database (Extended Data Fig.~\ref{fig:model}c).

{
We test the accuracy of the primary endogenic thermal model by running an inversion (described below) on a no-bombardment scenario, constrained by a curated database of \arar{} ages from  H~chondrites that are interpreted to have experienced minimally perturbed cooling histories~\cite{Trieloff2003}. 
The overall agreement between the posterior distributions and predicted H~parent body properties validates our radiogenic heating and conductive cooling model (Supplementary Fig.~3).
}

We then superimpose a bombardment reheating history over the primary cooling history by simulating impacts that reheat a fixed fractional volume of each shell within the body.
To simplify our model and underlying code, we assume that impacts only deposit energy and neither excavate nor implant material beyond the original body volume, in line with the probabilistic nature of our model asteroid.

Although the distribution of impact-heating in asteroidal bodies is heterogenous and complex \cite{Davison2012postimpact,Bland2014pressure}, both the H and LL chondrites record evidence for impacts while the body was still cooling from radiogenic heating \cite{Edwards2020accretion,Goudy2023evidence}, and H chondrites directly record the presence of nearly kilometer-scale melt sheets \cite{Rubin1983nature} that would have promoted heat transfer into parent body interiors. 
Indeed, suprasolidus type~7 and impact-melt chondrites are byproducts of impact heating in early asteroid planetesimals and are observed in every chondrite family evaluated in this study (Table~\ref{tab:ages}). 
Based on this reasoning, we conclude that collisions in the early solar system were capable of depositing significant amounts of thermal energy into chondritic planetesimal interiors.  

\ImpactChron{} accommodates three different geometries for the zone reheated by an impact: a cone, a parabololoid, or a hemisphere.
Since the shape, depth, and radius of the reheating zone beneath an impact site would vary drastically with impactor size and trajectory, we do not assume a single finite morphology for reheating.
Instead, we simulate a more probabilistic reheating scenario that heats a cone extending to the center of the body, allowing for reheating of interior zones that might be exposed by larger collisions.
We assign the outermost radius of the cone to cover 1\% of the asteroid's circumference to limit geometric error stemming from our simplified assumption of disk-shaped reheating volumes (which do not account for curvature) for each shell.
We justify this assumption in a later section.

We model the bombardment history of the model asteroid with one or more fluxes of impactors that reheat fractions of the body as described above.
Each bombardment event is modeled by an exponentially decaying flux defined by an initial impact flux $F_o$ (impacts/My), $e$-folding timescale $\tau$ (My), and bombardment onset date $t_o$ (M\yss{}).
The bombardment flux for any given date $t$ in M\yss{} is
\begin{equation}
    F =
\begin{cases}
 F_o \cdot exp\left(-\frac{t-t_o}{\tau}\right), & \text{if } t \ge t_o\\
 0, & \text{if } t<t_o
\end{cases}
\end{equation}\label{eq:flux}
This formulation results in fractional impacts (Extended Data Fig.~\ref{fig:model}d), which we accept given the approximate nature of our modeled volumes of impact reheating.
At each timestep in the model, we sum all fluxes $F$ to calculate the total number of ``impacts'' ($n_t = \sum F_t \cdot \Delta t$, where the model timestep is $\Delta t=1$~My) and scale the volumetric proportion of each shell reheated per impact  ($v^z_t$) by $n_t$.

We assume complete resetting of the \kar{} system within the reheated zone and instantaneous cooling through Ar closure.
{While we consider this assumption sufficient for the relatively low-temperature \kar{} system, this does not afford us the ability to simultaneously model the differently temperature-dependent behavior of multiple thermochronometers, hence we only consider the \kar{} system.}
For each shell in the body at each timestep, we reset equal proportions (scaled by $n_t \cdot v^z_t$) of the cooling age(s) recorded within that shell to the timestep age.
Thus, for the first timestep of an impact flux ($t_1$), $n_{t1} \cdot v^z_{t1}$ of the primary cooling age ($t_o$) is reset to the age of that timestep ($t_1$).
For the second timestep of that impact flux ($t_2$), $\frac{1}{2} \cdot n_{t2} \cdot v^z_{t2}$ of $t_o$ and $t_1$ are set to an age of $t_2$. 
And so on.
When iterated over each timestep of the simulation, this produces a matrix of fractional volumes for each modeled shell of the asteroid, with rows and columns corresponding to time/age and radial depth in the parent body.
By summing the volumetric proportions corresponding to each timestep, we get a distribution of cooling ages for the model asteroid (Extended Data Fig.~\ref{fig:model}c).

\subsection{Bayesian inversion}
To reconstruct the planetary history that accounts for the observed distribution of \kar{} system ages, we employ a MCMC method that uses this distribution of measured ages as a prior.
The thermochronologic model described above returns a distribution of cooling ages that correspond to the chosen suite of parameters describing the environmental, chemical, physical and, material properties of a model asteroid as well as its bombardment history (Table~\ref{tab:params}).
To explore this parameter space and estimate posterior distributions for each parameter, we use a modification of the Metropolis algorithm \cite{Metropolis1953equation}, based on the underlying statistical architecture of \cite{Keller2018}.

In addition to the prior distribution of measured \kar{} system ages, we incorporate a comprehensive set of ``parameter priors,'' compiled from published data, to constrain all the parameters describing the enivronmental, cosmochemical, material, and asteroidal properties of the simulation (Table~\ref{tab:params}).
We parameterized each prior as a log-normal distribution based on either the shape of its distribution {(the histogram is approximately log-normal)} or the fact that the property must be $>0$ by definition (e.g.~temperatures in Kelvin).
{Using a log-normal distribution ensures that the MCMC cannot jump to physically impossible negative values for such properties.}  
For each prior, we compiled a variety of measurements or estimates of the parameter, which included discrete values, normal distributions, and uniform distributions for data reported as ranges.
We then calculated the natural-logarithm-space $\mu$ and $\sigma$ of each distribution from $10^6$ random samples of the compiled data.
The two exceptions to this approach are the very precise solar age (assigned to the oldest Ca-Al-rich inclusions), which we treat as a constant in our model, and the initial \iso{26}{Al}/\iso{27}{Al} of the protoplanetary disk which was reported as a normal distribution (Table~\ref{tab:params}). 

Since the bombardment history of the asteroid belt is poorly constrained, we do not have prior distributions for the initial impactor flux, $e$-folding time, or onset of each bombardment episode. 
We assume uniform distributions that span the entire simulation timescale (\unf{2000,4567.3}~Ma) for the bombardment onset and $e$-folding time (Table~\ref{tab:params}).
Since the volume of impact reheating is simulated nonphysically, setting a maximum on the number of impacts need only accommodate an upper-limit reheating scenario and prevent numerical instability at extreme values. 
Since our parameterization of the reheating zone would completely reheat the full volume of the sphere with $\sim$4000~impacts, we set an arbitrary upperbound of $10^4$~My$^{-1}$ (Table~\ref{tab:params}).

The model accommodates up to three bombardment events ($\alpha$, $\beta$, $\gamma$), each described by its own $t_o$, $F_o$, and $\tau$ (Table~\ref{tab:params}).
In simulations with multiple bombardment histories, we assign $\alpha$ as a primordial flux anchored to the solar age ($t_o=0$~M\yss{}), and the MCMC algorithm explores the value(s) of $t_o$ for the other flux(es) ($\beta$, $\gamma$) within the time domain. 
We impart two rules on these fluxes to ensure reproducible model behavior.
(1.) To avoid unnecessary bimodal distributions, we require each flux  to occur in chronological order ($t_o\alpha<t_o\beta<t_o\gamma$), which prevents the Markov chains from ``swapping'' parameter spaces.
(2.) We also require that all post-accretion fluxes ($\beta$, $\gamma$) have a shorter $e$-folding timescale than the primordial flux, framing bombardment events as transient episodes of enhanced asteroidal collisions that recover more quickly than the background rate of intersecting orbits.
Any steps that violate either of these rules are rejected.

Thus, for a given model result---a distribution of simulated \arar{} ages---we calculate the log-likelihood ($\ell$) that the collection of measured ages (with corresponding, normally distributed uncertainties) were drawn from the simulated distribution.
We add to this $\ell$ the log-likelihoods that each simulated parameter was drawn from its corresponding prior distribution.
For each step of the Markov chain, we randomly perturb one parameter at a time to explore the full parameter space and accept proposals for any given step $i$ with a probability of $min\{\ell_i - \ell_{i-1}, 1\}$. 
After each accepted step, we scale the step size of the most recently perturbed variable by a constant tuned for an acceptance rate of $\sim$50\%. 
After an extended burn-in/warm-up period ($8 \cdot 10^5$~steps), we record 10$^6$ subsequent steps of the Markov chain as the stationary distribution.

\subsection{Evaluating and testing assumptions}\label{sec:assume}
In this section, we consider some of the assumptions made to simplify our model {and construct the \kar{} age database.}

\subsubsection{Decay constants}
{
Above, we justify our decision to calibrate our database of \kar{} ages with the SJ77 decay constants.
Some studies have suggested that these constants underestimate early solar system ages (relative to the absolute U-Pb thermochronometric system) by $\ge$30~Ma \cite{Trieloff2003,Schwarz2011comment}.
However, more recent work on a rapidly quenched system indicates that SJ77-calculated \arar{} ages underestimate absolute Pb-Pb ages by $\lesssim$10~Ma~\cite{Turrin202340ar}.
Here, we explore how two alternative age calibrations affect the results from inversions of our preferred model (two bombardments---one primordial, one post-accretion).
First, we test the effects of linearly shifting all \arar{} ages in our database 10~Ma earlier in solar system history (Supplementary Fig.~4).
Second, we test the effects of using the more contemporary decay constants calibrated from high-precision geochronology of $<$3~Ga terrestrial standards~\cite{Renne2010joint,Renne2011response} (Supplementary Fig.~5).
}

{
In both cases, the inversion converges on posteriors with median bombardment onset dates $>$12~M\yss{} (Supplementary Figs.~4, 5), that is, younger than the median onset date yielded from the SJ77-calibrated database.
This is initially surprising, since both alternative age calibrations yield systematically older ages than those calculated with the SJ77 constants.
However, we emphasize that the MCMC does not fit individual ages, but rather a distribution of ages. 
Shifting the entire distribution to earlier timescales (whether linearly or non-linearly) establishes a new shape of distribution that is modelled by a complex, nonlinear endogenic/exogenic thermal model.
Thus, a shift in ages will not necessarily yield a proportional shift in the bombardment onset date.
In conclusion, while debate among appropriate decay constants and age calibrations for the \arar{} system add some uncertainty to our conclusions, the effects do not challenge our primary conclusion that GPM likely post-dated dissipation of the gaseous protoplanetary disk.
}

\subsubsection{Ar closure temperature} 
Since Ar loss from mineral crystal lattices is a diffusive process, effective Ar closure temperature is dependent on many variables, including mineralogy, grain size, and cooling rate. 
In result, empirically derived Ar closure temperatures for chondrites range broadly (300 -- 800~K) due to variability of these properties among samples.
Moreover, the concept of closure temperature itself is a fictitious (albeit useful) simplification of the concurrent processes of radiogenic production and temperature-dependent diffusive loss of atoms within a thermochronometric mineral system~\cite{Dodson1973}.
In highly constrained mineral systems, fully modeling the production-diffusion process can simulate accurate thermochronologic histories~\cite{Blackburn2017accretion,Edwards2020accretion}.
However, in the case of bulk chondrite Ar isotope measurements with few to no mineralogic constraints (the typical case for the ages used in this study), such complete diffusion paramterization is infeasible. 
To circumvent this heterogeneity and complexity, we instead treat chondrite Ar closure temperatures as distributions rather than absolute values and explore this parameter with the MCMC algorithm, allowing the Bayesian posterior to accommodate its inherent variability.
We assert that exploring an effective Ar closure temperature as a free parameter in our Bayesian inversion allows us to capture the heterogeneity and complexity of Ar production-diffusion in chondritic mineral systems without the computational overhead and parametric uncertainty of explicitly simulating Ar diffusion. 

\subsubsection{Geometry of impact heating} 
We assume a nonphysical geometry---a cone extending to the planetesimal center---to simulate impact reheating.
This assumption allows us to consider impact reheating in a more agnostic, probabilistic fashion by resetting all depths and petrologic types in equal proportions.
This should in part capture the thermal effects of impact reheating of material from the asteroid interior exposed by large, more catastrophic disruptions. 
Though we assume that the parent body was not catastrophically disrupted during the timescales of primary cooling (see below), the delivery of type~6 chondrites to Earth requires eventual excavation of deeper material.
Nonetheless, the consequences of full-radius reheating are relatively minor for the deepest portions of our model asteroids.
Since planetary centers cool slowest, they are still hot (i.e.~above the Ar closure temperature) and insensitive to thermochronologic resetting during early reheating events.
For instance, median conditions from our prefered model ($\tau \sim 20$~Ma) result in negligibly low fluxes ($<$1~\textperthousand{} of $F_o$) after 150~My, about half the time it takes for the asteroid center to cool below Ar closure after accretion ($>$300~My).
Correspondingly, our model ignores reheating events prior to thermochronologic closure (discussed further below).

To test whether our selected parameterization is consistent with more physical approaches, we compare the results of reheating a parabolic region of fixed depth corresponding to impactors of 15 and 1~km diameters in a 2-bombardment scenario (1 primordial, as in our preferred case, Fig.~\ref{fig:impacts}).
We approximate the relative dimensions of simulated reheating (per impactor) from the results of prior impact heating work~\cite{Davison2012postimpact}.
The results show that reheating to $\gg1$~km depths is necessary: a smaller impactor (1~km diameter) yields posterior parameter distributions that mimic a no-impact scenario (Supplementary Fig.~6, Extended Data Fig.~\ref{fig:0Fposts}).
For sufficient heating depths, posterior bombardment histories are similar between the partial- and full-radius impact-heating parameterizations:
the posterior distributions of bombardment parameters for a $\ge$15~km impactor diameter (results for a 20~km impactor, not shown, are nearly identical)  are similar to those produced by the full-radius reheating approach (compare Extended Data Fig.~8 to Supplementary Fig.~7 and Fig.~\ref{fig:impacts} to Supplementary Fig.~8).
Thus, as long as sufficiently deep (type~5--6) material is reheated by our simulated impact fluxes, the posteriors are largely unchanged.

\subsubsection{Two-stage thermal model} 
As described above, we use a two-stage thermal model: we first model the unperturbed cooling history of a model asteroid with an analytical solution and then superimpose a secondary impact-reheating history over the primary cooling history.
While using this analytical solution vastly reduces computation time, it has a few drawbacks.
First, it requires constant material parameters (e.g. density, thermal conductivity), though these are known to vary as a function of temperature~\cite{Yomogida1983physical} and shock histories~\cite{Opeil2012stony}.
Additionally, we assume instantaneous accretion of the planetesimal to a fixed radius.
In each case, our use of the MCMC algorithm allows much (if not all) of the error stemming from these assumptions to be captured within the variation of the posterior estimates for each parameter. 

Our use of the two-stage model requires the assumption that the OC, RC, and EC parent planetesimals were not catastrophically disrupted prior to body-wide cooling below Ar closure temperatures. 
This assumption that planetesimals did not experience early disruptions is supported by several lines of evidence: thermochronologic evidence for the survival of OCs to $\ge$60--90~M\yss{}~\cite{Blackburn2017accretion,Edwards2020accretion} and observational/dynamical evidence for the long-term preservation of large asteroidal bodies~\cite{Bottke2005fossilized} (diameters $>$120~km, compare to diameters of $\sim$300~km from Table~1).
Using the median values of the posterior estimates of our preferred model (Fig.~\ref{fig:impacts}, Extended Data Fig.~\ref{fig:2Fposts}), the planetesimal center cools through Ar closure after $\sim$300~My, but 50\% (by volume) of the same body cools through Ar closure within the first 100~M\yss{}. 
We conclude that parent bodies of the chondrites examined in this study were likely not catastrophically disrupted within the most important timescales of the model and that any samples from the deep interior that were prematurely quenched by disruption reflect a relatively minor volume of material. 
Nonetheless, this is a significant assumption in our model that future studies could ameliorate by employing more physical thermal histories.

Finally, the two-stage thermal model structure ignores all impact events prior to the primary cooling age of any given shell.
The logic behind our assumption is simple: if the shell had not already cooled through closure, the addition of energy by an impact would not reset a not-yet closed thermochronologic system. 
Instead, this energy might prolong a cooling history, which is accounted for (at least partly) by subsequent impacts in the flux following the primary closure time of the corresponding shell.
The drawback of this approach is that there is little statistical sensitivity to simulated impact scenarios that begin early in time with very large fluxes, prior to widespread closure of the parent body.
However, this drawback reflects an inherent limitation of thermochronology: persistent high temperatures (either sustained or reheated) erase previous thermal information.
Thus, similar to nature, our model effectively ignores early, short-lived bombardment events.
Since the primary cooling ages of the outermost shells of our model asteroids are typically $\ge$5~M\yss{}, bombardment onsets prior to this timestep are not well-constrained. 
Thus, we suspect that the exaggerated early tail of bombardment onset dates near 0~M\yss{} in Fig.~\ref{fig:impacts} may be an artifact of this insensitivity.

\subsubsection{Age of the solar system}

{The architecture of our code requires a fixed solar time-zero ($t_{ss}$=0~M\yss), and we use the earliest-formed CAIs as our datum.
In our simulations, we use the widely applied datum of 4567.3~Ma, calculated from n=4 CAI samples~\cite{Connelly2012absolute}. 
However, another study precisely estimates the earliest episode of CAI formation at 4568.2~Ma~\cite{Bouvier2010age}, and recent cosmochronological models estimate an absolute age of 4568.7~Ma for the eldest CAIs~\cite{Piralla2023unified}.
While any of these ages are defensible as a solar datum, the results of our simulations are insensitive to variability in solar system age.
Using our inversion with an upperbound estimate for the age of CAIs ($t_{ss}$=4568.7~Ma) yields posterior distributions (Supplementary Fig.~9) that appear indistinguishable from those corresponding to $t_{ss}$=4567.3~Ma (Fig.~\ref{fig:impacts}).
The median and mean bombardment onset ages for the older datum scenario are both shifted (relative to the younger datum scenario) approximately 0.9~My later, which is similar to the 1.4~My difference between the two solar data (the 0.5~My discrepancy is $<$3\% of the range of the 50\% confidence interval).}

\section*{Data Availability}
All \kar{} system ages examined in this study and/or used in statistical codes are tabulated in Supplementary Table 1, and corresponding literature sources are tabulated in Supplementary Table 2. 
The data used to estimate priors in Table~\ref{tab:params} are available in the literature sources referenced therein. Detailed calculations of priors in Table~\ref{tab:params} and recalibrated \kar{} system ages are available at \url{https://github.com/grahamedwards/ImpactChron.jl/tree/main/data} and archived at \url{https://doi.org/10.5281/zenodo.11163986}.
Posterior distributions may be reproduced using the \href{https://github.com/grahamedwards/ImpactChron.jl}{\texttt{ImpactChron.jl}} software package.

\section*{Code Availability} 
All code used to analyze data and perform Markov chain Monte Carlo algorithms are available at \url{https://github.com/grahamedwards/ImpactChron.jl} and archived at \url{https://doi.org/10.5281/zenodo.11163986}.

\section*{Acknowledgements}
We are indebted to Maggie A.~Thompson for many insightful conversations and thoughtful feedback on an early draft of this manuscript. 
We thank Munazza K.~Alam for foundational discussions about giant planet migration mechanisms.
We thank Cyril P.~Opeil, S.J.~for sharing his profound insight into chondrite thermal properties. 
We thank the three reviewers of this manuscript for their careful revisions that substantially improved its quality and Dr.~Luca Maltagliati for careful editorial handling.
G.H.E. was supported by NSF Award \#2102591. 
C.W.S. was supported by an Undergraduate Research Assistantship at Dartmouth (Spring 2022).

\section*{Author Contributions}
G.H.E. ran simulations and wrote the manuscript. 
G.H.E. and C.B.K. wrote the code and interpreted results. 
G.H.E., C.B.K., and E.R.N. conceived of the study. 
G.H.E. and C.W.S. compiled the thermochronologic age database and wrote age recalculation codes. 
All authors contributed to editing the manuscript.

\section*{Competing interests}
The authors declare no competing interests.

\clearpage

\begin{table}

    \caption{
    Parameters in the asteroid thermochronologic code and prior distributions used in the Bayesian inversion.
    Priors are either a constant, a uniform distribution \unf{a,b}, a normal distribution $\mu\pm\sigma$, or a lognormal distribution \lognorm{\mu,\sigma}. Log-normal parameters are reported in log-space with linear-space approximations.
    M\yss{} denotes My after the solar age and $t_{max}$ the model time domain upper limit.
    } \label{tab:params}
    
    \begin{tabular}{l l l l r}
    & \multicolumn{2}{l}{\bfseries Parameter} & \bfseries Prior & \bfseries Reference \\
    \toprule 
    \multicolumn{5}{l}{\bfseries Environmental}  \\
    & $t_{ss}$ & Solar age (oldest CAIs) & 4567.3~Ma & \cite{Connelly2012absolute}\\
    & $T_m$ & Midplane temperature & \lognorm{5.4,0.5} $\sim$ 210 K & \cite{Woolum1999astronomical}\\
    \multicolumn{5}{l}{\bfseries Cosmochemical}\\
    &  $^{26}Al_o$ & Initial \textsuperscript{26}Al/\textsuperscript{27}Al & $5.23 \pm 0.06 \times 10^{-5}$& \cite{Jacobsen200826al} \\
    & $[Al]$ & Al abundance & \lognorm{-4.6, 0.1} $\sim$ 1.0~wt\,\% & \cite{Lodders1998planetary}\\
    & $T_c$ & Ar closure temperature & \lognorm{6.2,0.3} $\sim490$~K & \cite{Turner1978early,Bogard2009arar,Weirich2009arar,Bogard2010arar,Trieloff2003} \\
    \multicolumn{5}{l}{\bfseries Asteroid}  \\
    &  $R$ & Radius & \lognorm{11.9,0.2} $\sim$ 150 km& \cite{Trieloff2003,Trieloff2022evolution,Henke2013thermal,Blackburn2017accretion,Gail2019thermal, Edwards2020accretion}\\
    & $t_a$ & Time of accretion & \lognorm{0.70,0.08} $\sim$ 2.0 M\yss{} & \cite{Trieloff2003,Trieloff2022evolution,Henke2013thermal,Blackburn2017accretion,Gail2019thermal, Edwards2020accretion,Sugiura2014correlated} \\
    \multicolumn{5}{l}{\bfseries Material}  \\
    & $\rho$ & Bulk density & \lognorm{8.12,0.04} $\sim 3400~\frac{kg}{m^{3}}$ & \cite{Flynn2018physical,Macke2010survey}\\
    & $C_p$ & Specific heat capacity & \lognorm{6.73,0.08} $\sim 850~\frac{J}{kg \cdot K}$ & \cite{Wach2013specific} \\
    & $K$ & Thermal conductivity & \lognorm{0.3,0.6} $\sim 1.4~\frac{W}{m \cdot K}$ & \cite{Yomogida1983physical,Opeil2010thermal,Opeil2012stony} \\
    \multicolumn{5}{l}{\bfseries Bombardment}  \\
    & $t_o$ & Start time 	& \unf{0,t_{max}} (M\yss{}) &  \\
    & $F_o$ & Initial flux  	& \unf{0,10^4} (My$^{-1}$) &  \\
    & $\tau$ & $e$-folding time & \unf{0,t_{max}} (My) &  \\
      \bottomrule
      \\
      \end{tabular}
    
    \end{table}

    
    \begin{table}
        \caption{
        Population statistics of $>$2000~Ma \arar{} dates (n=97), used as priors in this study.
        Each cell reports the counts of all samples affiliated with the corresponding chondrite group (columns) and petrologic type (rows). 
        We combine type~7 and impact-melted chondrites given their shared suprasolidus histories.
        This table excludes (n=2) R chondrites and (n=2) ungrouped E-type impact melts.
        The summed totals exceed the sample size, since individual meteorites may have multiple group or petrologic type affiliations (e.g.~regolith breccias). 
        }
        \label{tab:ages}
        \centering
        \begin{tabular}{>{\bfseries}r | c c c c c | c}
         \multicolumn{2}{c}{} & \multicolumn{3}{c}{\textbf{Group}} & \multicolumn{2}{c}{} \\
        \textbf{Petrologic Type} & \textbf{H} & \textbf{L} & \textbf{LL} & \textbf{EH} & \textbf{EL} & \textbf{Total} \\
        \toprule
            Type 3 & 2 & 4 & 4 & 2 & 1 & 13 \\
            Type 4 & 8 & 5 & 3 & 3 & 0 & 19 \\
            Type 5 & 14 & 1 & 6 & 2 & 1 & 24 \\
            Type 6 & 12 & 4 & 6 & 0 & 10 & 32 \\
            Type 7 / melt & 3 & 3 & 2 & 5 & 2 & 15 \\
            \midrule
            Total & 39 & 17 & 21 & 12 & 14 &  
        \end{tabular}
    
    \end{table}

    \clearpage
\graphicspath{{figs/}}


\renewcommand{\figurename}{Extended Data Figure}
\setcounter{figure}{0}

\renewcommand{\tablename}{Extended Data Table}
\setcounter{table}{0}

\begin{figure}
    \centering
    \includegraphics[width=.9\textwidth,keepaspectratio]{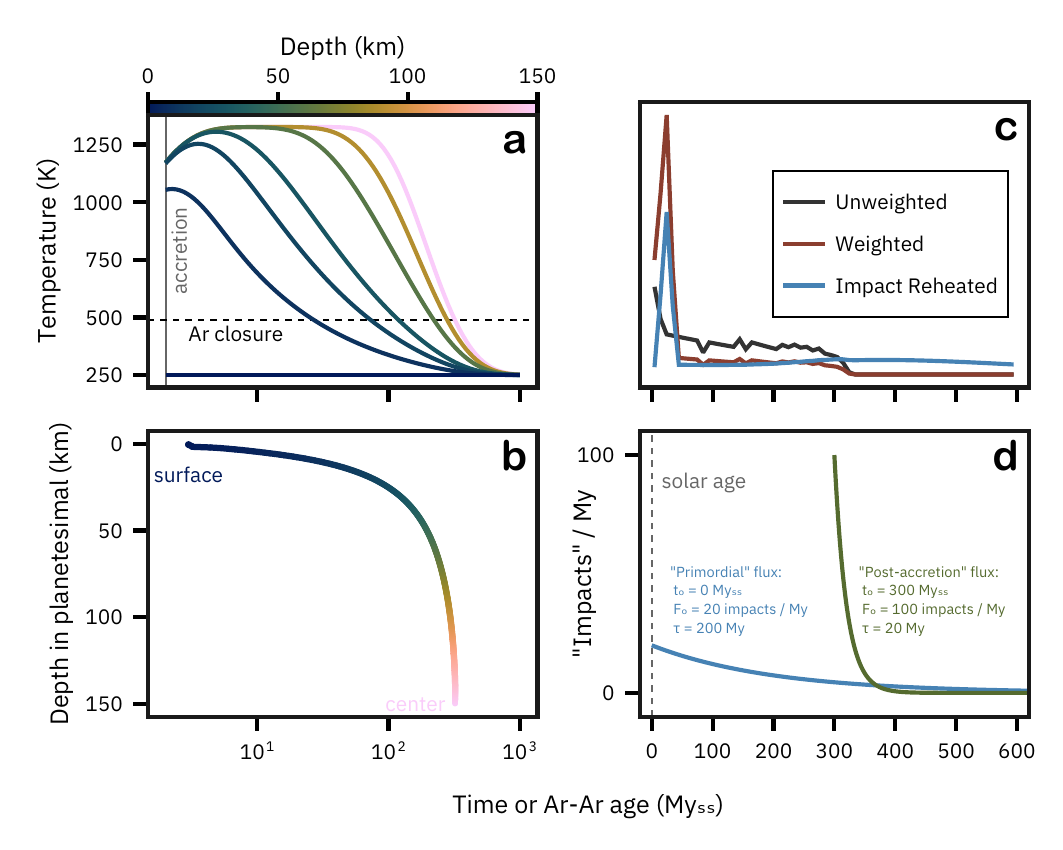}
    \caption{\textbf{Workflow of asteroidal-scale thermochronologic model.}
    M\yss{} denotes My after solar system formation. 
    (\textbf{a}) We use an analytical solution to the heat equation in a radiogenically heated, conductively cooling spherical body \cite{Hevey2006model,Carslaw1959conduction}.
    Each curve traces the time-temperature history at a depth in the simulated asteroid after instantaneous accretion (solid grey line). 
    As the temperature of a given depth passes below the effective closure temperature of Ar (dashed black line), we assign that timestep as the \arar{} age (panel \textbf{b}).
    Panels \textbf{a} and \textbf{b} share color scales and a logarithmic timescale.
    The parameters used in these simulations are the priors' central tendencies in Table~\ref{tab:params}.
    (\textbf{c}) We calculate a distribution of \arar{} cooling ages from the calculated ages and volumetric proportions of each simulated radial shell (black curve, labeled ``Unweighted"). 
    We assign a petrologic type to each depth in the body based on the peak temperature of its time-temperature history (\textbf{a}) and recalculate a petrologic type-weighted distribution of ages (``Weighted", red curve). 
    This step increases the proportion of early ages from shallower depths.
    The blue curve (``Impact Reheated'') depicts the effect of impact reheating by the primordial impact flux depicted in panel \textbf{d}. 
    (\textbf{d}) For simulations with bombardment histories, we ``reheat'' the body at a range of depths with one or more exponentially decaying fluxes of impacts. 
    Panel \textbf{d} depicts two such fluxes: a ``primordial'' flux anchored to the solar age (0 M\yss{}) and a ``post-accretion'' flux beginning 300~M\yss{}.
    The primordial flux has a lower initial flux ($20$~My$^{-1}$) and longer $e$-folding timescale ($200$~My), resulting in a mild/protracted bombardment.
    The post-accretion flux has a higher initial flux ($100$~My$^{-1}$) and shorter $e$-folding timescale ($20$~My), resulting in an intense/brief bombardment.
    }
    \label{fig:model}
\end{figure}


\begin{figure}
    \centering
    \includegraphics[width=\textwidth,keepaspectratio]{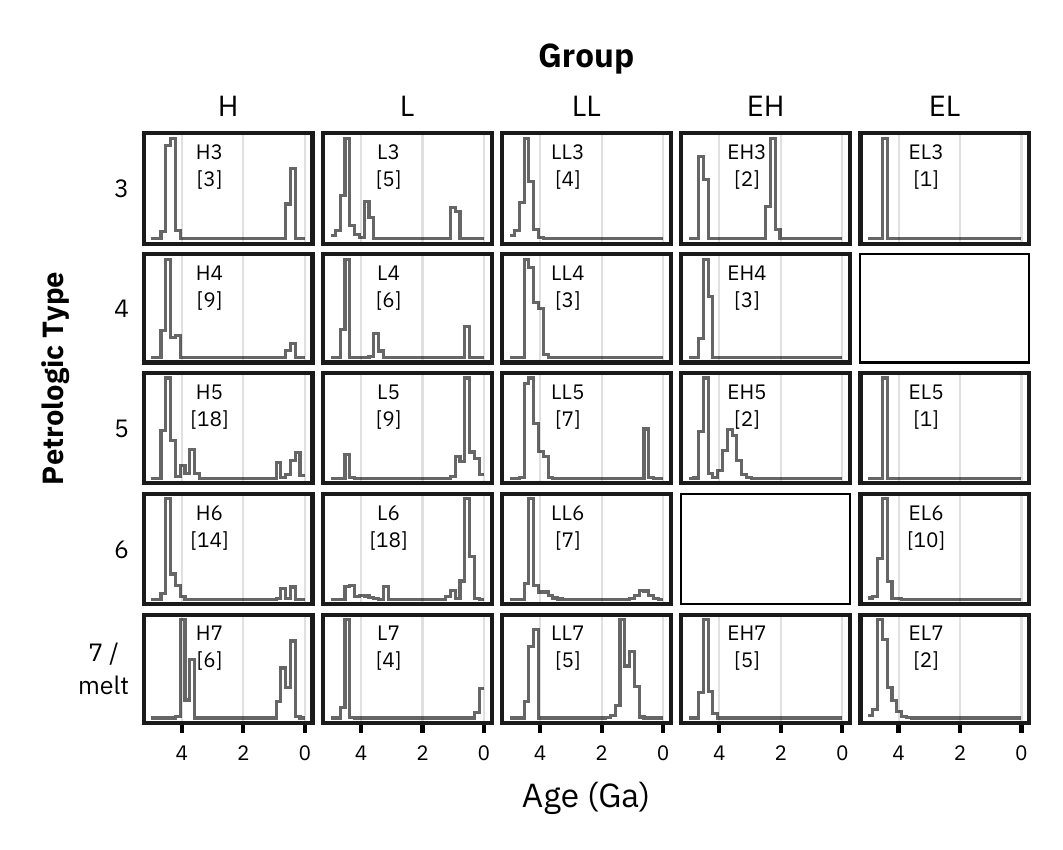}
    \caption{\textbf{
    Distributions of \arar{} cooling ages compiled for this study (n=136).}  
    Histograms of summed age distributions arranged as a grid of pairings for each group (H, L, LL, EH, EL in columns) and petrologic type (3--7 in rows).
    Within each panel, bracketed numbers indicate sample size, and empty panels (EH6, EL4) indicate no samples of that classification.
    As in Table~\ref{tab:ages}, some meteorites occur in multiple panels due to nonexclusive classifications, and we combine type~7s and impact-melted (``melt'') chondrites given their shared suprasolidus histories.
    These plots exclude the ages of Rumuruti (R3--6, $4460\pm 11$~Ma), Acfer~217 (R3--5, $4300 \pm 70$~Ma), and ungrouped E-type impact melts Zak\l{}odzie ($4503 \pm 9$~Ma) and QUE~97348 ($4444 \pm 17$~Ma). 
    While there are few systematic trends across petrologic types, there are some trends within groups.
    L~chondrites, particularly L5 and L6 types, have a large proportion of $<$1~Ga ages. 
    EH and EL chondrite \arar{} ages largely overlap and are typically $>$4~Ga. 
    All ages and affiliations are tabulated in Supplementary Table 1. 
    }
    \label{fig:agegrid}
\end{figure}

\begin{figure}
    \centering
    \includegraphics[width=\textwidth, keepaspectratio]{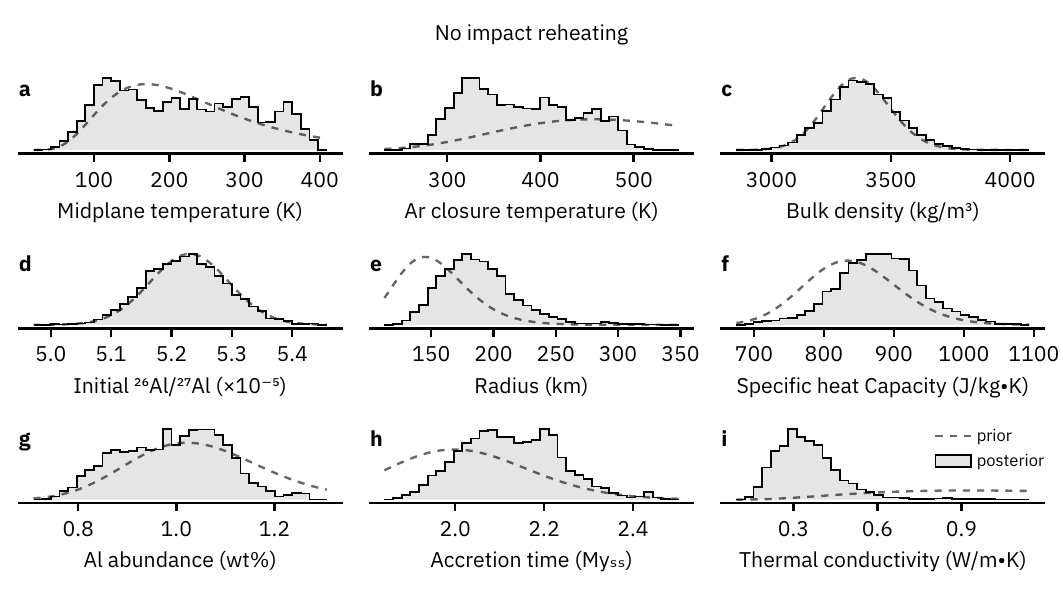}
    \caption{
    \textbf{Posterior distributions of thermochronologic model parameters for a simulated asteroid with no impact reheating.}
    Each histogram reflects $10^6$ steps of post-burn-in MCMC simulation. 
    Dashed lines demarcate prior distributions of each corresponding parameter. 
    Posterior distributions of bulk density (\textbf{c}) and initial \iso{26}{Al}/\iso{27}{Al} (\textbf{d}) agree well with their priors, while all other parameters appear inconsistent with their respective priors.
    Extended Data Table~\ref{tab:posts01} reports summary statistics.
    }
    \label{fig:0Fposts}
\end{figure}


\begin{figure}
    \centering
    \includegraphics[width=\textwidth, keepaspectratio]{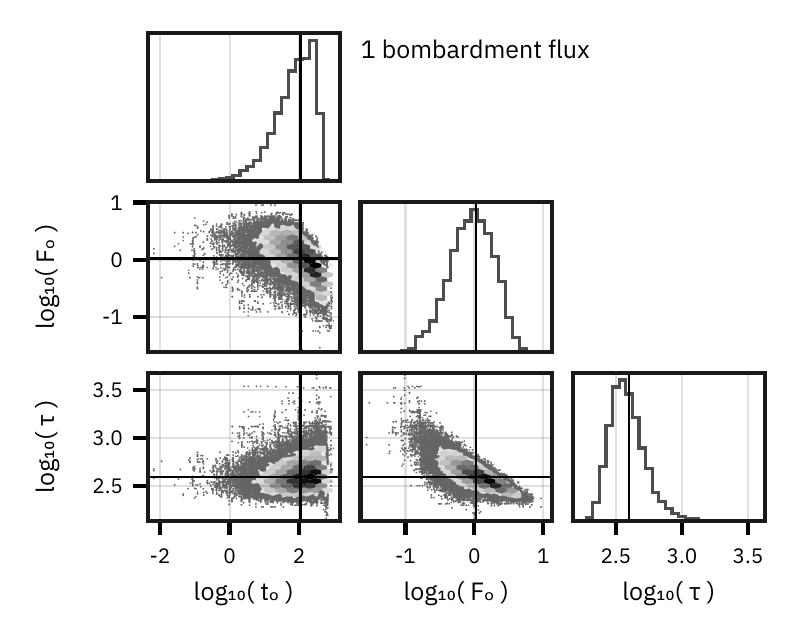}
    \caption{
    \textbf{Corner plot diagram for the bombardment parameters of a single bombardment scenario.}
    Each distribution reflects 10$^6$ Markov chain steps and black lines trace the median values ($M$) of each parameter.
    The initial impactor flux ($F_o,~M \sim 1$~My$^{-1}$) decreases as the onset date ($t_o,~M \sim 100$~M\yss{}) and $e$-folding decay timescale ($\tau,~M \sim 400$~My) increase. 
    See Fig.~\ref{fig:impacts} caption for description of corner plot layout.
    Extended Data Table~\ref{tab:posts01} reports summary statistics. 
    }
    \label{fig:1Fcorner}
\end{figure}


\begin{figure}
    \centering
    \includegraphics[width=\textwidth, keepaspectratio]{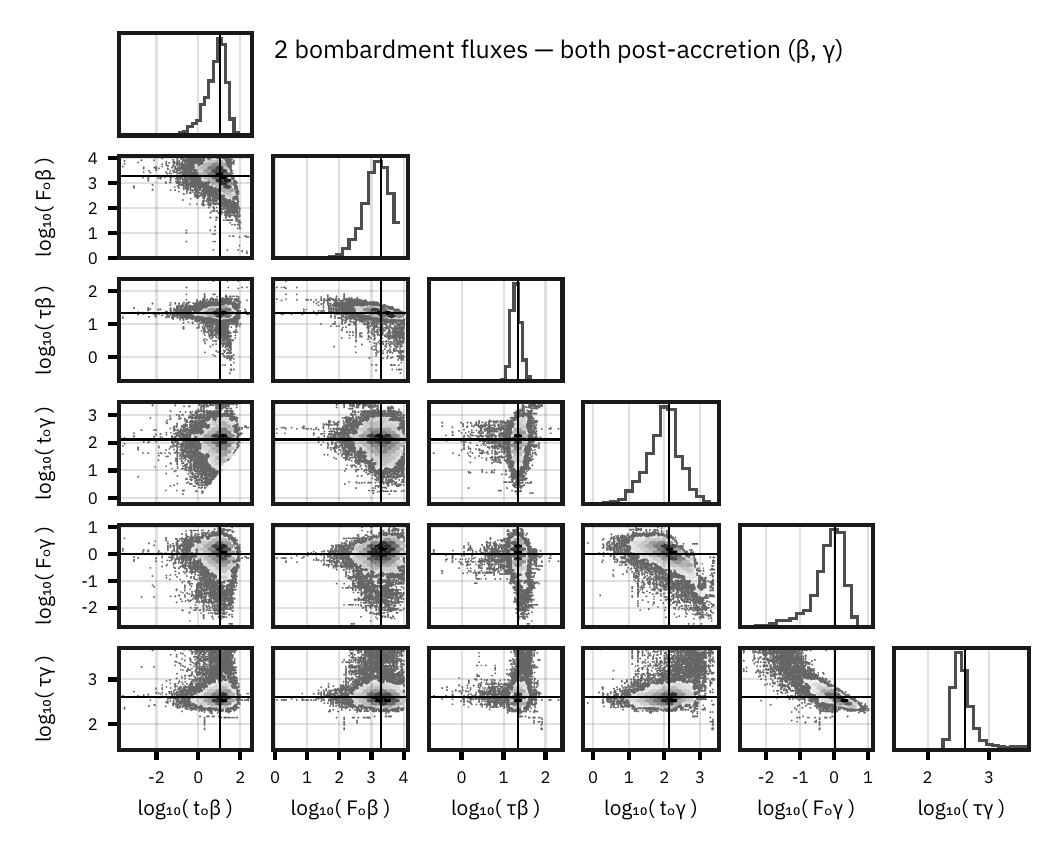}
    \caption{
    \textbf{Corner plot diagram for the bombardment parameters of a 2-flux history.}
    The scenario is similar to that reported in Fig.~\ref{fig:impacts}, except that both bombardments ($\beta$, $\gamma$) are ``post-accretion'' events with onset dates ($t_o$) that are explored as free parameters by the MCMC algorithm. 
    Each distribution reflects 10$^6$ Markov chain steps and black lines trace the median values ($M$) of each parameter.
    Impactor flux $\beta$ begins with a median onset date of $M(t_o\beta) \sim 10$~M\yss{} and is intense/brief ($M(F_o\beta) > 1000$~My$^{-1}$, $M(\tau\beta) \sim 10$~My).
    Impactor flux $\gamma$ begins far later ($M(t_o\gamma) \sim 100$~M\yss{}), but is mild/protracted ($M(F_o\gamma) \sim 1$~My$^{-1}$, $M(\tau\gamma) \sim 500$~My), similar to the primordial flux in Fig.~\ref{fig:impacts}b. 
    There is little apparent correlation between fluxes, but within each flux, $F_o$ scales inversely with longer $\tau$ and later $t_o$.
    See Fig.~\ref{fig:impacts} caption for description of corner plot layout.
    Extended Data Table~\ref{tab:posts23} reports summary statistics.
    }
    \label{fig:2Fcorner}
\end{figure}


\begin{figure}
    \centering
    \includegraphics[width=\textwidth, keepaspectratio]{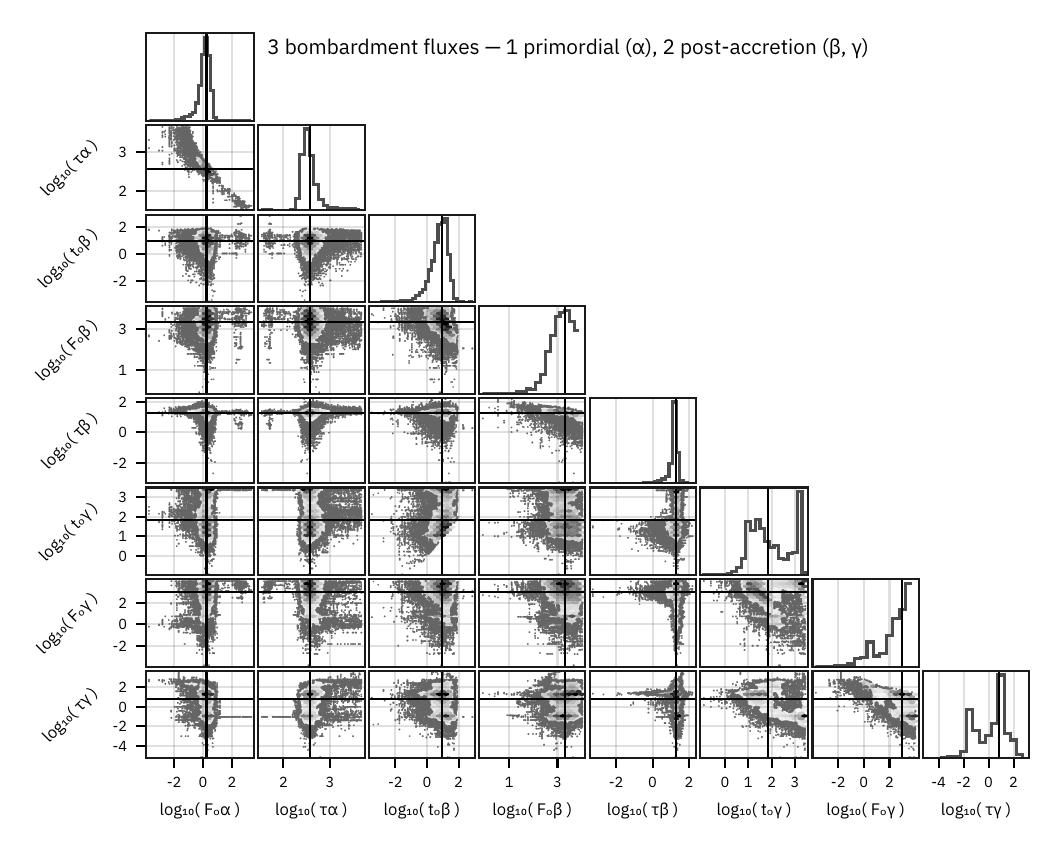}
    \caption{
    \textbf{Corner plot diagram for the bombardment parameters of a 3-flux history.}
    Bombardment $\alpha$ is a ``primordial'' flux ($t_o$=0~M\yss{}), whereas the MCMC algorithm explores values of $t_o$ for bombardments $\beta$ and $\gamma$.
    Each distribution reflects 10$^6$ Markov chain steps and black lines trace the median values ($M$) of each parameter.
    Bombardment events $\alpha$ and $\beta$ show a similar pattern to that observed in 2-bombardment simulations (Fig.~\ref{fig:impacts}, Extended Data Fig.~\ref{fig:2Fcorner}).
    The primordial impactor flux ($\alpha$) is mild/protracted ($M(F_o\alpha) \sim 1$~My$^{-1}$, $M(\tau\alpha) \sim 400$~My). 
    Impactor flux $\beta$ is intense/brief ($M(F_o\beta) > 1000$~My$^{-1}$, $M(\tau\beta) \sim 10$~My), with a median onset date of $\sim $10~M\yss{}.
    Impactor flux $\gamma$ exhibits a bimodal distribution in $t_o$ and $\tau$: an early mode ($t_o\gamma < 100$~M\yss{}) with $\tau \sim10$~My  and a later mode ($t_o\gamma >1000$~M\yss{}) with $\tau \sim 0.1$~My.
    The initial impact flux is unimodally large ($M(F_o\beta) > 1000$~My$^{-1}$). 
    There is little apparent correlation between fluxes, but within each flux, $F_o$ scales inversely with longer $\tau$ and later $t_o$.
    Extended Data Table~\ref{tab:posts23} reports summary statistics.
    }
    \label{fig:3Fcorner}
\end{figure}


\begin{figure}
    \centering
    \includegraphics[width=\textwidth, keepaspectratio]{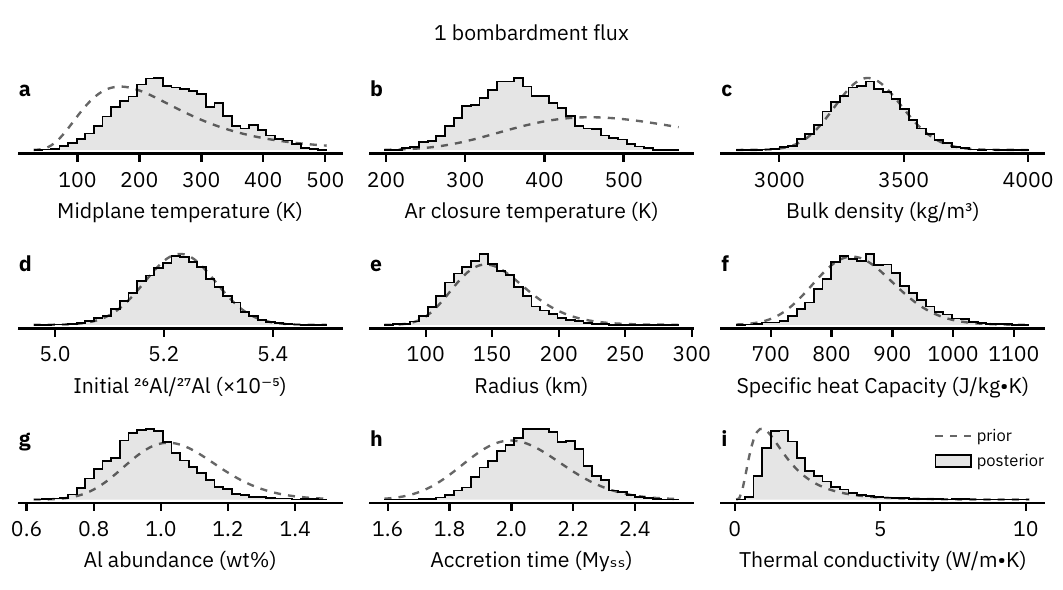}
    \caption{
    \textbf{Posterior distributions of non-bombardment thermochronologic model parameters for a simulated asteroid with a single bombardment flux.}
    Each histogram reflects $10^6$ steps of post-burn-in MCMC simulation. 
    Dashed lines demarcate prior distributions of each corresponding parameter. 
    Posterior distributions of bulk density (\textbf{c}), initial \iso{26}{Al}/\iso{27}{Al} (\textbf{d}), asteroid radius (\textbf{e}), and specific heat capacity (\textbf{f}) agree well with their priors, while all other parameters appear inconsistent with their respective priors.
    Extended Data Table~\ref{tab:posts01} reports summary statistics.
    Extended Data Fig.~\ref{fig:1Fcorner} depicts bombardment parameter posteriors.
    }
    \label{fig:1Fposts}
\end{figure}


\begin{figure}
    \centering
    \includegraphics[width=\textwidth, keepaspectratio]{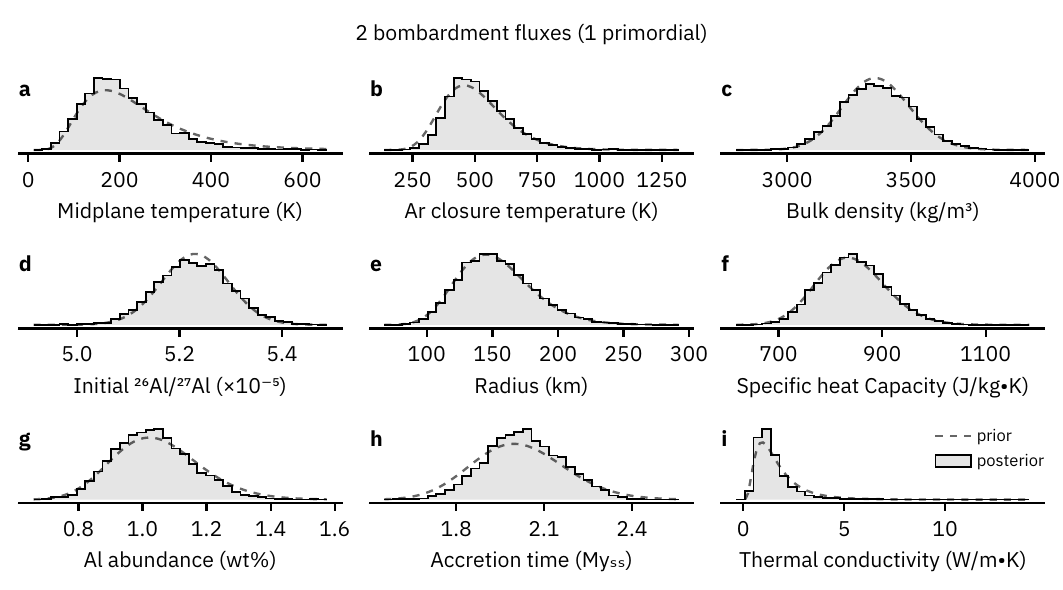}
    \caption{
    \textbf{Posterior distributions of non-bombardment thermochronologic model parameters for a simulated asteroid with two bombardment fluxes.}
    These posteriors are from the Markov chain simulation depicted in Figs.~\ref{fig:fits}c and \ref{fig:impacts}.
    Each histogram reflects $10^6$ steps of post-burn-in MCMC simulation. 
    Dashed lines demarcate prior distributions of each corresponding parameter.
    All posterior distributions agree well with their priors.
    All simulations with 2--3 bombardment histories (Figs.~\ref{fig:fits}c--d, Extended Data Figs.~\ref{fig:2Fcorner}, \ref{fig:3Fcorner}) have nearly identical prior-posterior relationships for non-bombardment parameters (Supplementary Figs.~1,2).
    Extended Data Table~\ref{tab:posts23} reports summary statistics. 
    }
    \label{fig:2Fposts}
\end{figure}

\clearpage



\begin{table}

\caption{
\textbf{Summary statistics of parameter posterior distributions for no-impact and single bombardment simulations.}
Parameter names (``Param.") correspond to Table~\ref{tab:params}. 
Priors are either a constant, a uniform distribution \unf{a,b}, a normal distribution $\mathcal{N}(\mu,\sigma)$, or a lognormal distribution \lognorm{\mu,\sigma}. 
For log-normally distributed parameters, means and standard deviations ($\mu \pm \sigma$) are calculated from and reported as natural logarithms. 
Median and 95\% credible interval ($M \pm CI$) are all reported in linear-space. 
}
\label{tab:posts01}
\scriptsize
\begin{tabular}{l c|c|cc|cc}
&&& \multicolumn{2}{c|}{No impact flux} & \multicolumn{2}{c}{1 bombardment}  \\ 
Param. & Units & Prior & $\mu \pm \sigma$ & $M \pm CI$ & $\mu \pm \sigma$ & $M \pm CI$ \\
\toprule 
$t_{ss}$ & $Ma$ & $4567.3$ & $ 4567.3 $ & $ 4567.3 $ & $ 4567.3 $ & $ 4567.3 $ \\
$T_m$ & $K$ & $log\mathcal{N}(5.3517, 0.4691)$ & $5.28 \pm 0.46$ & $209 ^{+165}_{-131}$ & $5.49 \pm 0.34$ & $250 ^{+174}_{-136}$ \\
$^{26}Al_o$ & $\times 10^{-5}$ & $\mathcal{N}(5.23, 0.65)$ & $5.23 \pm 0.07$ & $5.22 ^{+0.13}_{-0.12}$ & $5.23 \pm 0.07$ & $5.23 ^{+0.13}_{-0.13}$ \\
$[Al]$ & $wt\,\%$ & $log\mathcal{N}(-4.5665, 0.1316)$ & $-4.62 \pm 0.11$ & $0.999 ^{+0.187}_{-0.199}$ & $-4.64 \pm 0.12$ & $0.966 ^{+0.2499}_{-0.194}$ \\
$T_c$ & $K$ & $log\mathcal{N}(6.1914, 0.2571)$ & $5.92 \pm 0.15$ & $371 ^{+113}_{-84}$ & $5.9 \pm 0.2$ & $366 ^{+127}_{-102}$ \\
$R$ & $km$ & $log\mathcal{N}(11.92, 0.1869)$ & $12.1 \pm 0.2$ & $184 ^{+69}_{-45}$ & $11.9 \pm 0.2$ & $145 ^{+58}_{-42}$ \\
$t_a$ & $My_{ss}$ & $log\mathcal{N}(0.6983, 0.0792)$ & $0.753 \pm 0.051$ & $2.12 ^{+0.23}_{-0.19}$ & $0.735 \pm 0.06004$ & $2.09 ^{+0.24}_{-0.24}$ \\
$\rho$ & $kg/m^3$ & $log\mathcal{N}(8.1199, 0.03996)$ & $8.13 \pm 0.04$ & $3380 ^{+300}_{-270}$ & $8.12 \pm 0.04$ & $3360 ^{+290}_{-260}$ \\
$C_p$ & $J/(kg\,K)$ & $log\mathcal{N}(6.732, 0.07899)$ & $6.78 \pm 0.07$ & $879 ^{+122}_{-125}$ & $6.75 \pm 0.07$ & $856 ^{+140}_{-112}$ \\
$K$ & $W/(m\,K)$ & $log\mathcal{N}(0.3319, 0.6335)$ & $-1.1 \pm 0.3$ & $0.335 ^{+0.277}_{-0.149}$ & $0.627 \pm 0.452$ & $1.83 ^{+2.91}_{-1.02}$ \\
$t_o^\alpha$ & $My_{ss}$ & $\mathcal{U}[0.0, 2567.3]$ & --- & --- & $159 \pm 148$ & $109 ^{+448}_{-106}$ \\
$\tau^\alpha$ & $My$ & $\mathcal{U}[0.0, 4567.3]$ & --- & --- & $434 \pm 169$ & $396 ^{+448}_{-145}$ \\
$F_o^\alpha$ & $My^{-1}$ & $\mathcal{U}[0.0, 10000.0]$ & --- & --- & $1.29 \pm 0.92$ & $1.05 ^{+2.59}_{-0.83}$ \\
\bottomrule
\end{tabular}
\end{table}

\begin{sidewaystable}
\caption{
\textbf{Summary statistics of parameter posterior distributions for 2- and 3-bombardment simulations.}
Parameter names (``Param.") correspond to Table~\ref{tab:params}. 
Priors are either a constant, a uniform distribution \unf{a,b}, a normal distribution $\mathcal{N}(\mu,\sigma)$, or a lognormal distribution \lognorm{\mu,\sigma}. 
For log-normally distributed parameters, means and standard deviations ($\mu \pm \sigma$) are calculated from and reported as natural logarithms. 
Median and 95\% credible interval ($M \pm CI$) are all reported in linear-space. 
}
\label{tab:posts23}
\scriptsize
    \begin{tabular}{lc|c|cc|cc|cc}
&&&
\multicolumn{2}{c|}{2 bombardments $^\dag$} & 
\multicolumn{2}{c|}{2 bombardments $\ddag$} & 
\multicolumn{2}{c}{3 bombardments $^\S$} \\ 
Param. & Units & Prior & 
$\mu \pm \sigma$ & $M \pm CI$ & 
$\mu \pm \sigma$ & $M \pm CI$ & 
$\mu \pm \sigma$ & $M \pm CI$ \\
\toprule 
$t_{ss}$ & $Ma$ & $4567.3$ & $ 4567.3 $ & $ 4567.3 $ & $ 4567.3 $ & $ 4567.3 $ & $ 4567.3 $ & $ 4567.3 $ \\
$T_m$ & $K$ & $log\mathcal{N}(5.3517, 0.4691)$ & $5.26 \pm 0.41$ & $194 ^{+222}_{-113}$ & $5.22 \pm 0.42$ & $190 ^{+212}_{-115}$ & $5.25 \pm 0.42$ & $194 ^{+226}_{-114}$ \\
$^{26}Al_o$ & $\times 10^{-5}$ & $\mathcal{N}(5.23, 0.65)$ & $5.23 \pm 0.07$ & $5.23 ^{+0.14}_{-0.14}$ & $5.23 \pm 0.07$ & $5.23 ^{+0.14}_{-0.14}$ & $5.23 \pm 0.07$ & $5.23 ^{+0.13}_{-0.13}$ \\
$[Al]$ & $wt\,\%$ & $log\mathcal{N}(-4.5665, 0.1316)$ & $-4.58 \pm 0.12$ & $1.03 ^{+0.26}_{-0.21}$ & $-4.57 \pm 0.12$ & $1.04 ^{+0.26}_{-0.21}$ & $-4.58 \pm 0.11$ & $1.03 ^{+0.24}_{-0.2}$ \\
$T_c$ & $K$ & $log\mathcal{N}(6.1914, 0.2571)$ & $6.22 \pm 0.23$ & $499 ^{+287}_{-171}$ & $6.22 \pm 0.24$ & $494 ^{+319}_{-168}$ & $6.24 \pm 0.23$ & $508 ^{+310}_{-180}$ \\
$R$ & $km$ & $log\mathcal{N}(11.92, 0.1869)$ & $11.9 \pm 0.2$ & $151 ^{+64}_{-45}$ & $11.9 \pm 0.2$ & $154 ^{+70}_{-46}$ & $12 \pm 0$ & $155 ^{+68}_{-47}$ \\
$t_a$ & $My_{ss}$ & $log\mathcal{N}(0.6983, 0.0792)$ & $0.707 \pm 0.064$ & $2.03 ^{+0.26}_{-0.24}$ & $0.704 \pm 0.065$ & $2.03 ^{+0.26}_{-0.25}$ & $0.701 \pm 0.062$ & $2.02 ^{+0.26}_{-0.23}$ \\
$\rho$ & $kg/m^3$ & $log\mathcal{N}(8.1199, 0.03996)$ & $8.12 \pm 0.04$ & $3360 ^{+290}_{-260}$ & $8.12 \pm 0.04$ & $3360 ^{+290}_{-280}$ & $8.12 \pm 0.04$ & $3360 ^{+280}_{-270}$ \\
$C_p$ & $J/(kg\,K)$ & $log\mathcal{N}(6.732, 0.07899)$ & $6.74 \pm 0.08$ & $843 ^{+135}_{-114}$ & $6.74 \pm 0.08$ & $842 ^{+145}_{-119}$ & $6.74 \pm 0.08$ & $847 ^{+132}_{-122}$ \\
$K$ & $W/(m\,K)$ & $log\mathcal{N}(0.3319, 0.6335)$ & $0.234 \pm 0.5295$ & $1.23 ^{+2.75}_{-0.76}$ & $0.103 \pm 0.574$ & $1.11 ^{+2.396}_{-0.76}$ & $0.22 \pm 0.53$ & $1.23 ^{+2.58}_{-0.78}$ \\
$t_o^\alpha$ & $My_{ss}$ & $\mathcal{U}[0.0, 2567.3]$ & $ 0 $ & $ 0 $ & --- & --- & $ 0 $ & $ 0 $ \\
$\tau^\alpha$ & $My$ & $\mathcal{U}[0.0, 4567.3]$ & $440 \pm 344$ & $373 ^{+621}_{-133}$ & --- & --- & $484 \pm 458$ & $376 ^{+1369}_{-144}$ \\
$F_o^\alpha$ & $My^{-1}$ & $\mathcal{U}[0.0, 10000.0]$ & $2.11 \pm 1.39$ & $1.82 ^{+3.77}_{-1.59}$ & --- & --- & $4.74 \pm 39.31$ & $1.75 ^{+4.05}_{-1.68}$ \\
$t_o^\beta$ & $My_{ss}$ & $\mathcal{U}[0.0, 2567.3]$ & $15 \pm 14$ & $11.3 ^{+44.4}_{-10.98}$ & $13.4 \pm 12.1$ & $10.4 ^{+34.8}_{-10.0}$ & $13.2 \pm 13.8$ & $9.16 ^{+42.26}_{-8.86}$ \\
$\tau^\beta$ & $My$ & $\mathcal{U}[0.0, 4567.3]$ & $21.6 \pm 6.4$ & $21.2 ^{+13.8}_{-15.8}$ & $23 \pm 7$ & $22 ^{+17}_{-10}$ & $20.2 \pm 9.2$ & $20 ^{+20}_{-18}$ \\
$F_o^\beta$ & $My^{-1}$ & $\mathcal{U}[0.0, 10000.0]$ & $2460 \pm 2110$ & $1760 ^{+6560}_{-1580}$ & $2560 \pm 2090$ & $1930 ^{+6400}_{-1730}$ & $2820 \pm 2470$ & $2010 ^{+7160}_{-1870}$ \\
$t_o^\gamma$ & $My_{ss}$ & $\mathcal{U}[0.0, 2567.3]$ & --- & --- & $204 \pm 259$ & $131 ^{+781}_{-119}$ & $557 \pm 834$ & $70.8 ^{+2328.0}_{-64.8}$ \\
$\tau^\gamma$ & $My$ & $\mathcal{U}[0.0, 4567.3]$ & --- & --- & $569 \pm 608$ & $402 ^{+2386}_{-158}$ & $29.4 \pm 85.8$ & $6.46 ^{+235.76}_{-6.44}$ \\
$F_o^\gamma$ & $My^{-1}$ & $\mathcal{U}[0.0, 10000.0]$ & --- & --- & $1.28 \pm 1.04$ & $1.04 ^{+2.87}_{-1.01}$ & $2330 \pm 2890$ & $890 ^{+8495}_{-889}$ \\
\bottomrule 
\multicolumn{9}{l}{$^\dag$~One primordial, one post-accretion. Corresponds to Figs.~\ref{fig:fits}c,\ref{fig:impacts} and Extended Data Fig.~\ref{fig:2Fposts}.}\\
\multicolumn{9}{l}{$^\ddag$~Both post-accretion. Corresponds to Extended Data Fig.~\ref{fig:2Fcorner} and Supplementary Fig.~1. }\\
\multicolumn{9}{l}{$^\S$~One primordial, two post-accretion. Corresponds to Figs.~\ref{fig:fits}d, Extended Data Fig.~\ref{fig:3Fcorner}, and Supplementary Fig.~S2.}
\end{tabular}
\end{sidewaystable}

\end{document}